%% file: main-tr.tex
\documentclass[sigconf]{acmart}
\AtBeginDocument{%
  }

\copyrightyear{2025}
\acmYear{2025}
\acmConference[KDD '25]{Proceedings of the 31st ACM SIGKDD Conference on
Knowledge Discovery and Data Mining V.2}{August 3--7, 2025}{Toronto, ON,
Canada}
\acmBooktitle{Proceedings of the 31st ACM SIGKDD Conference on Knowledge
Discovery and Data Mining V.2 (KDD '25), August 3--7, 2025, Toronto, ON,
Canada}
\acmDOI{10.1145/3711896.3737012}
\acmISBN{979-8-4007-1454-2/2025/08}

\input{cmd}

\begin{document}

\title{\sketrag: A Cost-Efficient Multi-Granular Indexing Framework for \graphrag}

\author{Yiqian Huang}
\orcid{0009-0001-5601-3439}
\affiliation{%
  \institution{National University of Singapore}
  \country{Singapore}
}
\email{yiqian@comp.nus.edu.sg}

\author{Shiqi Zhang}
\authornote{Corresponding author.}
\orcid{0000-0002-7155-9579}
\affiliation{%
  \institution{National University of Singapore}
  \country{Singapore}
}
\affiliation{%
  \institution{PyroWis AI}
  \country{Singapore}
}
\email{shiqi@pyrowis.ai}

\author{Xiaokui Xiao}
\orcid{0000-0003-0914-4580}
\affiliation{%
  \institution{National University of Singapore}
    \country{Singapore}
}
\email{xkxiao@nus.edu.sg}

\renewcommand{\shortauthors}{Yiqian Huang, Shiqi Zhang, \& Xiaokui Xiao}

\input{sections/abs}




\maketitle

\input{sections/intro}
\input{sections/prelim}

\input{sections/related}

\input{sections/overview}

\input{sections/details}

\input{sections/exp}
\input{sections/conclusion}

\bibliographystyle{ACM-Reference-Format}
\bibliography{ref}

\end{document}

%% file: cmd.tex
\usepackage{xspace,balance,tabularx,multirow}
\usepackage{flushend}
\usepackage{tikz}
\usepackage{pgfplots}
\usetikzlibrary{patterns}
\pgfplotsset{compat=1.16}
\usepackage[ruled, vlined, linesnumbered]{algorithm2e}
\usepackage[noend]{algpseudocode}
\usepackage{xcolor}
\usepackage{float}
\usepackage{colortbl}
\usepackage{hyperref}
\usepackage{bbold}
\usepackage{textcomp}
\usepackage{pifont}
\SetKwComment{Comment}{$\triangleright$\ }{}
\usepackage{enumitem}
\usepackage{xcolor,framed}
\usepackage[aboveskip=-4pt]{subcaption}
\usepackage[normalem]{ulem}
\usepackage{tikz-3dplot}
\useunder{\uline}{\ul}{}

\newcommand{\revise}[1]{{#1}}

\newcommand{\argmaxk}[1]{\underset{#1}{\operatorname{arg}\,\operatorname{max-k}}\;}
\newcommand{\argmink}[1]{\underset{#1}{\operatorname{arg}\,\operatorname{min-k}}\;}
\newcommand{\eat}[1]{} 

\makeatletter
\newcommand*\bigcdot{\mathpalette\bigcdot@{.5}}
\newcommand*\bigcdot@[2]{\mathbin{\vcenter{\hbox{\scalebox{#2}{$\m@th#1\bullet$}}}}}
\makeatother

\SetKw{Break}{break}

\newcommand{\ie}{{i.e.},\xspace}
\newcommand{\eg}{{e.g.},\xspace}

\newcommand{\stitle}[1]{\noindent{\bf #1.\/}}

\newcommand{\G}{\mathcal{G}\xspace}
\newcommand{\W}{\mathcal{W}\xspace}

\newcommand{\N}{\mathcal{N}\xspace}
\newcommand{\Sset}{\mathcal{S}\xspace}

\newcommand{\Tset}{\mathcal{T}\xspace}
\newcommand{\temb}[1]{\boldsymbol{\upphi}(#1)}
\newcommand{\V}{\mathcal{V}\xspace}

\newcommand{\E}{\mathcal{E}\xspace}

\newcommand{\clen}{\ell\xspace}

\newcommand{\textgraph}{TAG}
\newcommand{\textrag}{\textsf{Text-RAG}\xspace}
\newcommand{\graphrag}{\textsf{Graph-RAG}\xspace}
\newcommand{\msgraphrag}{\textsf{MS-Graph-RAG}\xspace}
\newcommand{\kgrag}{\textsf{KG-RAG}\xspace}
\newcommand{\hybridrag}{\textsf{Hybrid-RAG}\xspace}
\newcommand{\hyde}{\textsf{HyDE}\xspace}
\newcommand{\lightrag}{\textsf{LightRAG}\xspace}
\newcommand{\lightragl}{\textsf{LightRAG-Local}\xspace}
\newcommand{\lightragg}{\textsf{LightRAG-Global}\xspace}
\newcommand{\lightragh}{\textsf{LightRAG-Hybrid}\xspace}
\newcommand{\hipporag}{\textsf{HippoRAG}\xspace}
\newcommand{\knnrag}{\textsf{KNNG-RAG}\xspace}
\newcommand{\keyrag}{\textsf{Keyword-RAG}\xspace}
\newcommand{\skeletonragu}{\textsf{Skeleton-RAG-U}\xspace}
\newcommand{\skeletonrag}{\textsf{Skeleton-RAG}\xspace}

\newcommand{\graphindex}{\textsf{KG-Index}\xspace}
\newcommand{\graphretrieval}{\textsf{KG-Retrieval}\xspace}
\newcommand{\sketrag}{\textsf{KET-RAG}\xspace}
\newcommand{\sketragu}{\textsf{KET-RAG-U}\xspace}
\newcommand{\sketragp}{\textsf{KET-RAG-P}\xspace}
\newcommand{\sketindex}{\texttt{KET-Index}\xspace}
\newcommand{\sketretrieval}{\texttt{KET-Retrieval}\xspace}

\definecolor{Red}{HTML}{E81123}
\definecolor{Orange}{HTML}{FF8C00}
\definecolor{Green}{HTML}{009E49}
\definecolor{LightBlue}{HTML}{00BCF2}
\definecolor{DeepBlue}{HTML}{001BA3}
\definecolor{Pink}{HTML}{F2028F}

\newenvironment{customlegend}[1][]{%
    \begingroup
    \csname pgfplots@init@cleared@structures\endcsname
    \pgfplotsset{#1}%
}{%
    \csname pgfplots@createlegend\endcsname
    \endgroup
}%

\def\addlegendimage{\csname pgfplots@addlegendimage\endcsname}

%% file: sections/abs.tex
\begin{abstract}
\textsf{Graph-RAG} constructs a knowledge graph from text chunks to improve retrieval in Large Language Model (LLM)-based question answering. It is particularly useful in domains such as biomedicine, law, and political science, where retrieval often requires multi-hop reasoning over proprietary documents. Some existing \textsf{Graph-RAG} systems construct KNN graphs based on text chunk relevance, but this coarse-grained approach fails to capture entity relationships within texts, leading to sub-par retrieval and generation quality. To address this, recent solutions leverage LLMs to extract entities and relationships from text chunks, constructing triplet-based knowledge graphs. However, this approach incurs significant indexing costs, especially for large document collections.

To ensure a good result accuracy while reducing the indexing cost, we propose \textsf{KET-RAG}, a multi-granular indexing framework. \textsf{KET-RAG} first identifies a small set of key text chunks and leverages an LLM to construct a knowledge graph skeleton. It then builds a text-keyword bipartite graph from all text chunks, serving as a lightweight alternative to a full knowledge graph. During retrieval, \textsf{KET-RAG} searches both structures: it follows the local search strategy of existing \textsf{Graph-RAG} systems on the skeleton while mimicking this search on the bipartite graph to improve retrieval quality. \revise{We evaluate 13 solutions on three real-world datasets}, demonstrating that \textsf{KET-RAG} outperforms all competitors in indexing cost, retrieval effectiveness, and generation quality. Notably, it achieves comparable or superior retrieval quality to Microsoft's \textsf{Graph-RAG} while reducing indexing costs by over an order of magnitude. Additionally, it improves the generation quality by up to 32.4\% while lowering indexing costs by around 20\%.
\end{abstract}

%% file: sections/intro.tex
\section{Introduction}
Given a set of text chunks $\Tset$, Graph-based Retrieval-Augmented Generation (\graphrag)~\cite{nebularag,edge2024local} enhances generative model inference by structuring $\Tset$ into a Text-Attributed Graph (\textgraph{}) $\G$ and retrieving relevant information from it.
Compared to \textrag~\cite{lewis2020retrieval}, which retrieves independent text chunks from $\Tset$, \graphrag captures relationships within and across text snippets to enhance multi-hop reasoning~\cite{peng2024graph,delile2024graph,jin2024graph}. \graphrag has gained widespread adoption across domains such as e-commerce~\cite{wang2022rete,xu2024retrieval}, biomedical research~\cite{delile2024graph,li2024dalk}, healthcare~\cite{chen24rarebench}, political science~\cite{mou2024unifying}, legal applications~\cite{colombo2024leveraging,kalra2024hypa}, and many others~\cite{alhanahnah2024depesrag,peng2024connecting}.

Some studies~\cite{li2024graph,wang2024knowledge} instantiate \textgraph{} $\G$ as a K-nearest-neighbor (KNN) graph, where nodes represent text chunks in $\Tset$, and edges encode semantic similarity or relevance. This approach maintains a low graph construction cost comparable to \textrag. However, it fails to capture entities and their relationships within text chunks, limiting retrieval effectiveness and degrading the quality of generated answers. 
To address this limitation, recent studies~\cite{li2024dalk,delile2024graph,edge2024local,gutierrez2024hipporag} have turned to triplet-based knowledge graphs, leveraging Large Language Models (LLMs) to extract structured (entity, relation, entity) triplets from text. This approach, known as \kgrag, enbales the LLM to filter out noise in raw documents and construct a more structured and interpretable knowledge base, significantly improving retrieval and generation quality. As a result, it has gained significant traction by major companies, including Microsoft~\cite{edge2024local}, Ant Group~\cite{antgrouprag}, Neo4j~\cite{neo4jrag}, and NebulaGraph~\cite{nebularag}.
However, \kgrag comes with a high indexing cost, particularly for large datasets. Even with the cost-efficient GPT-4o-mini API, processing a 3.2MB sample of the HotpotQA dataset~\cite{yang2018hotpotqa} costs \$21. 
In real-world applications, textual data often spans gigabytes to terabytes, making indexing costs prohibitively expensive. For instance, processing a single 5GB legal case~\cite{arnold2022ediscovery} incurs an estimated \$33K, posing a significant challenge for large-scale adoption.

To improve retrieval and generation quality while lowering indexing costs, we propose \sketrag, a cost-efficient multi-granular indexing framework for \graphrag. It comprises two key components: a knowledge graph skeleton and a text-keyword bipartite graph.
Instead of fully materializing the knowledge graph, \sketrag first identifies a set of core text chunks from $\Tset$ based on their PageRank centralities~\cite{page1999pagerank} in an intermediate KNN graph. It then constructs a skeleton of the complete knowledge graph using \kgrag described above. To prevent information loss from relying solely on this skeleton, \sketrag also builds a text-keyword bipartite from $\Tset$, serving as a lightweight alternative to \kgrag. By linking keywords to the text chunks in which they appear, keywords and their neighboring text chunks can be regarded as candidate entities and corresponding relations in the knowledge graph. During retrieval, \sketrag adopts the local search strategy of existing solutions but, unlike previous methods, extracts ego networks from both entity and keyword channels to facilitate LLM-based generation.

In experiments, we evaluate \revise{13 }solutions on three datasets across three key aspects: indexing cost, retrieval quality, and generation quality. Notably, \sketrag achieves retrieval quality comparable to or better than Microsoft's \graphrag~\cite{edge2024local}, the state-of-the-art \kgrag solution, while reducing indexing costs by over an order of magnitude. At the same time, it improves generation quality by up to 32.4\% while lowering indexing costs by approximately 20\%.
Furthermore, the core components of \sketrag, \skeletonrag and \keyrag, also function as effective stand-alone RAG solutions, balancing efficiency and quality. In particular, \skeletonrag reduces indexing costs by 20\% while maintaining retrieval quality, showing only minor performance drops in low-cost settings and achieving parity or even slight improvements in high-accuracy configurations. Meanwhile, \keyrag consistently outperforms the vanilla \textrag in both retrieval and generation quality, achieving up to 92.4\%, 133.3\%, 118.5\%, and 12.9\% relative improvements in Coverage, EM, F1 scores, and BERTScore, respectively, \revise{while exhibiting no compromise in retrieval latency.}

To summarize, we make the following contributions in this work:
\begin{itemize}[topsep=2pt,itemsep=1pt,parsep=0pt,partopsep=0pt,leftmargin=11pt]
    \item We propose \sketrag, a cost-efficient multi-granular indexing framework for \graphrag, integrating two complementary components to balance indexing cost and result quality.
    \item We introduce \skeletonrag, which constructs a knowledge graph skeleton by selecting core text chunks and leveraging LLMs to extract structured knowledge.
    \item We develop \keyrag, a lightweight text-keyword bipartite graph that mimics the retrieval paradigm of \kgrag while significantly reducing indexing costs.
    \item We conduct extensive experiments demonstrating the improvements of our proposed solutions.
\end{itemize}

%% file: sections/prelim.tex
\section{Preliminaries}
We first define the key terminologies and notations in Section~\ref{sec:notations}, and present the objective of this work in Section~\ref{sec:rag}. Finally, we review the state-of-the-art Microsoft's \graphrag in Section~\ref{sec:ms-graphrag}.

\subsection{Terminologies and Notations}\label{sec:notations}
Let $\Tset$ denote a set of text chunks, which is preprocessed from a set of documents. For simplicity, we assume that each text chunk $t_i \in \Tset$ is partitioned into chunks of equal length, denoted by $\clen$ tokens. The text embedding of each chunk $t_i$ is denoted as $\temb{t_i}$.
We define a \textit{text-attributed graph} (\textgraph) index as $\G = (\V, \E)$, where $\V$ and $\E$ are the sets of nodes and edges, respectively. 
In $\V$, each node is represented as $v_i=(t_i,\temb{t_i})$, where $t_i$ is the textual attribute and $\temb{t_i}$ is its text embedding. Analogously, each edge $e_{i,j}\in \E$ between nodes $v_i$ and $v_j$ is denoted as $e_{i,j}=(v_i, v_j, t_{i,j},\temb{t_{i,j}})$ if it is text-attributed; otherwise, it is simply $e_{i,j} = (v_i, v_j)$. In addition, we use calligraphic uppercase letters (e.g., $\mathcal{\Sset}$) to denote sets of nodes or edges. For text information, $\temb{\cdot}$ represents the text embedding function, the function $\bigoplus(\Sset)$ represents the concatenation of all text in $\Sset$, and $\left|\bigoplus(\Sset)\right|$ denotes the token count of the concatenated string.
The frequently-used notations are summarized in Table~\ref{tab:notation}.

\begin{table}[!t]
\centering
\renewcommand{\arraystretch}{1.1}
\begin{small}
\caption{Frequently used notations.}\vspace{-2.4mm} \label{tab:notation}
\begin{tabular}{rp{2.2in}}	
    \toprule
    \bf Notation & \bf Description \\
    \midrule
    $\Tset, t_i, \clen$ & Text chunk set $\Tset$ with each chunk $t_i$ in length $\clen$. \\
    $\temb{t_{i}}$ & The text embedding of text $t_i$.\\
    $\G = (\V, \E)$   &  Node set $\V$ and edge set $\E$ in \textgraph{} index $\G$. \\
    $\bigoplus(\Sset), \left|\bigoplus(\Sset)\right|$   &  the concatenated texts from $\Sset$ and its token count. \\
    $K, \beta, \tau$   &  the integer $K$ of KNN graph, the budget ratio $\beta$, and the number of splits $\tau$. \\
    $\lambda, \theta$   &  the context limit $\lambda$ and the retrieval ratio $\theta$. \\
    \bottomrule
\end{tabular}
\end{small}
\end{table}

\subsection{Problem Formulation}\label{sec:rag}
The Retrieval-Augmented Generation (RAG) framework consists of three main stages: indexing, retrieving, and generation. 

Given a set of text chunks $\Tset$, the indexing stage of \textrag~\cite{lewis2020retrieval} generates a text embedding $\temb{t_i}$ for each $t_i \in \Tset$. During the retrieval stage, \textrag computes the query embedding $\temb{q}$ and retrieves text chunks from $\Tset$ that are most similar to the query in the embedding space. Finally, the retrieved text chunks are incorporated into a predefined prompt for a large language model (LLM) to generate the final response.
In contrast, \graphrag constructs a \textgraph{} index $\G = (\V, \E)$ from $\Tset$ during indexing. Existing methods primarily build two types of $\G$: (i) a K-Nearest-Neighbors (KNN) graph (\knnrag), where each text chunk is a node, and edges represent similarity-based connections~\cite{li2024graph,wang2024knowledge}; or (ii) a knowledge graph (\kgrag), where LLMs extract (entity, relation, entity) triplets from text~\cite{li2024dalk,delile2024graph,edge2024local,gutierrez2024hipporag}.
During retrieval, \graphrag computes $\temb{q}$ and identifies \textit{seed nodes} in $\G$ that have the most similar text embeddings to the query. A subgraph is then extracted via local search, serialized into natural language, and incorporated into a predefined prompt for LLM-based response generation.

\stitle{Our objective} 
This work aims to develop a cost-efficient and effective approach for \graphrag to construct a \textgraph{} index $\G$ from $\Tset$, achieving lower indexing costs yet higher result accuracy in the widely adopted local search scenario~\cite{li2024dalk,delile2024graph,edge2024local,gutierrez2024hipporag}.

\subsection{Microsoft's \graphrag}\label{sec:ms-graphrag}
Microsoft proposed the \graphrag system~\cite{edge2024local}, which constructs a knowledge graph index with multi-level communities and employs tailored strategies for both local and global search. In this section, we focus on its indexing and retrieval operations for local search, which are relevant to our work.

Algorithm~\ref{alg:graphrag-index} outlines the pseudo-code for constructing the graph index $\G = (\V_e \cup \V_t, \E)$, where $\V_e$ and $\V_t$ represent entities and text chunks, respectively. Given a text chunk set $\Tset$, \graphindex first generates a text embedding for each $t_i \in \Tset$ and treats the corresponding text snippets as text-attributed nodes, forming the node set $\V_t$ (Line 1).  
For each node $v_i \in \V_t$, \graphindex leverages a predefined LLM to process $t_i$ in two steps: \textit{entity identification} and \textit{relationship extraction}, obtaining the sets of entities and relations ($\V_i$ and $\E_i$) (Line 3). These extracted entities and relationships are then added to $\V_e$ and $\E$, respectively (Line 4). 
For each extracted entity or relationship $x \in (\V_i \cup \E_i)$, the LLM generates a textual description $t_x$ along with its embedding $\temb{t_x}$. Additionally, \graphindex links each text chunk node $v_i$ to its corresponding extracted entities and relationships (Line 5).

\begin{algorithm}[!t]
\KwIn{The text chunk set $\Tset$.}
\KwOut{A \textgraph{} index $\G$.}
$\V_t \gets \{(t_i,\temb{t_i})\mid t_i\in \Tset\}$; ~ $\V_e \gets \emptyset; ~ \E \gets \emptyset$\;
\For{each $v_i= (t_i, \temb{t_i}) \in \V_t$}{
$\V_i,~\E_i \gets$ entities and edges extracted by LLM from $t_i$\;
$\V_e \gets \V_e \cup \V_i$;~$\E \gets \E \cup \E_i$\;
$\E \gets \E \cup \{(v_i,x) \mid x \in (\V_i \cup \E_i)\}$\;
}
\Return{$(\V_e\cup\V_t, \E)$;}
\caption{\graphindex $(\Tset)$}
\label{alg:graphrag-index}
\end{algorithm}

The local search procedure retrieves context from entities, relationships, and text chunks. Algorithm~\ref{alg:graphrag-retrieval} outlines the retrieval process, following the default context limit ratio across channels.  
Given a graph index $\G$ and a context limit $\lambda$, \graphretrieval retrieves contexts in the order of seed entities, relationships, and text chunks. Specifically, it first retrieves 10 entity nodes ($\Sset_e$) with embeddings most similar to the query embedding based on Euclidean distance (Line 1).  It then retrieves relationships ($\Sset_r$) until the combined token count from $\Sset_e\cup \Sset_r$ reaches $\lambda/2$ (Line 2). Relationships are prioritized based on adjacency to $\Sset_e$, with those connecting two seed entities ranking higher.  
For text chunk retrieval, \graphretrieval retrieves text chunks most adjacent to $\Sset_e$ and $\Sset_r$ until the total context length reaches $\lambda$ (Line 3). At last, the retrieved texts from $\Sset_e \cup \Sset_r \cup \Sset_t$ are concatenated to form the final context, which is returned as input for generation (Line 4).

%% file: sections/related.tex
\section{Related Works}\label{sec:related}
Beyond Microsoft's \graphrag, we review other indexing and retrieval approaches within existing \graphrag frameworks. For a comprehensive review, we refer readers to surveys~\cite{peng2024graph,fan24survey, zhou2025depth}.

\begin{algorithm}[!t]
\KwIn{A \textgraph{} index $\G=(\V_e\cup\V_t,\E)$, a query $q$, context length limit $\lambda$}
\KwOut{The context $C$} 
$\Sset_e \gets \argmink{v_i\in \V_e} \text{euc}(v_i,q)$, where $k=10$\;
$\Sset_r \gets \argmaxk{e_{i,j}\in \E} \text{adj}(\Sset_e,e_{i,j})$, s.t.\ ${\scriptstyle \left|\bigoplus(\Sset_r)\right|+\left|\bigoplus(\Sset_e)\right| = \lambda/2}$;\\
$\Sset_t \gets \argmaxk{v_{i}\in \V_t} \text{adj}(\Sset_e\cup\Sset_r,v_{i})$, s.t.\ ${\scriptstyle \left|\bigoplus(\Sset_t)\right| = \lambda/2}$;\\
\Return{$\bigoplus(\Sset_e\cup\Sset_r\cup\Sset_t)$;} 
\caption{\graphretrieval$(\G, q, \lambda)$}
\label{alg:graphrag-retrieval}
\end{algorithm}

Given $\Tset$, the core of constructing a KNN graph is measuring text chunk similarity. In particular, \citet{li2024graph} consider both structural and lexical similarities. Structural similarity is based on the physical adjacency of text chunks, linking neighboring passages in $\G$. Lexical similarity connects chunk nodes that share common keywords, which are extracted using LLM-based prompting.
\citet{wang2024knowledge} leverage multiple lexical similarity measures. Two chunk nodes are connected if they share keywords extracted using TF-IDF~\cite{ramos2003using}, contain common Wikipedia entities identified via TAGME~\cite{min2019knowledge}, or exhibit semantic similarity based on text embeddings.
Recent works~\cite{li2024dalk,delile2024graph,edge2024local,gutierrez2024hipporag} use LLMs to extract (entity, relation, entity) triplets from $\Tset$ to build knowledge graph indices, improving retrieval quality.
Akin to Algorithm~\ref{alg:graphrag-index}, 
\citet{delile2024graph} employ PubmedBERT~\cite{gu2021domain} to extract triplets from biomedical texts and link entities to the text chunks in which they appear. \citet{gutierrez2024hipporag} further enrich graph connectivity by linking semantically similar entities within the knowledge graph.
Several studies construct \textgraph{} indices using explicit relationships, such as co-authorship or citation links in academic papers~\cite{munikoti2023atlantic}, trade relationships between companies~\cite{cao24companykg}, and other structured connections~\cite{jin2024graph}. These publicly curated indices are beyond the scope of this work.
In summary, \knnrag is a more cost-effective solution but lacks fine-grained entity relationships. In contrast, \kgrag achieves higher effectiveness but incurs significant indexing costs, particularly for large $\Tset$.

To retrieve the most relevant subgraph given a query, various local search strategies have been proposed. Below, we focus on methods that utilize heuristic or traditional graph algorithms. 
Similar to Algorithm~\ref{alg:graphrag-retrieval}, \citet{jin2024graph} extract ego networks from seed nodes to enhance retrieval precision. In addition, \citet{li2024dalk} propose a two-step approach that first extracts a k-hop subgraph from seeds, followed by reranking and pruning the subgraph using LLMs. Other approaches include shortest path retrieval, where \citet{delile2024graph} and \citet{mavromatis2024gnn} retrieve the shortest path between seed nodes. \citet{gutierrez2024hipporag} use personalized PageRank to extract relevant subgraphs. \textsf{G-Retriever}~\cite{he2024g} focuses on query-aware subgraph generation by balancing semantic similarity to the query with subgraph generation cost. Hybrid retrieval methods, such as \hybridrag~\cite{sarmah2024hybridrag} and \textsf{EWEK-QA}~\cite{dehghan2024ewek}, enhance retrieval by querying both text and knowledge graphs.

%% file: sections/overview.tex
\begin{figure*}[!t]
    \centering
    \includegraphics[width=0.85\textwidth]{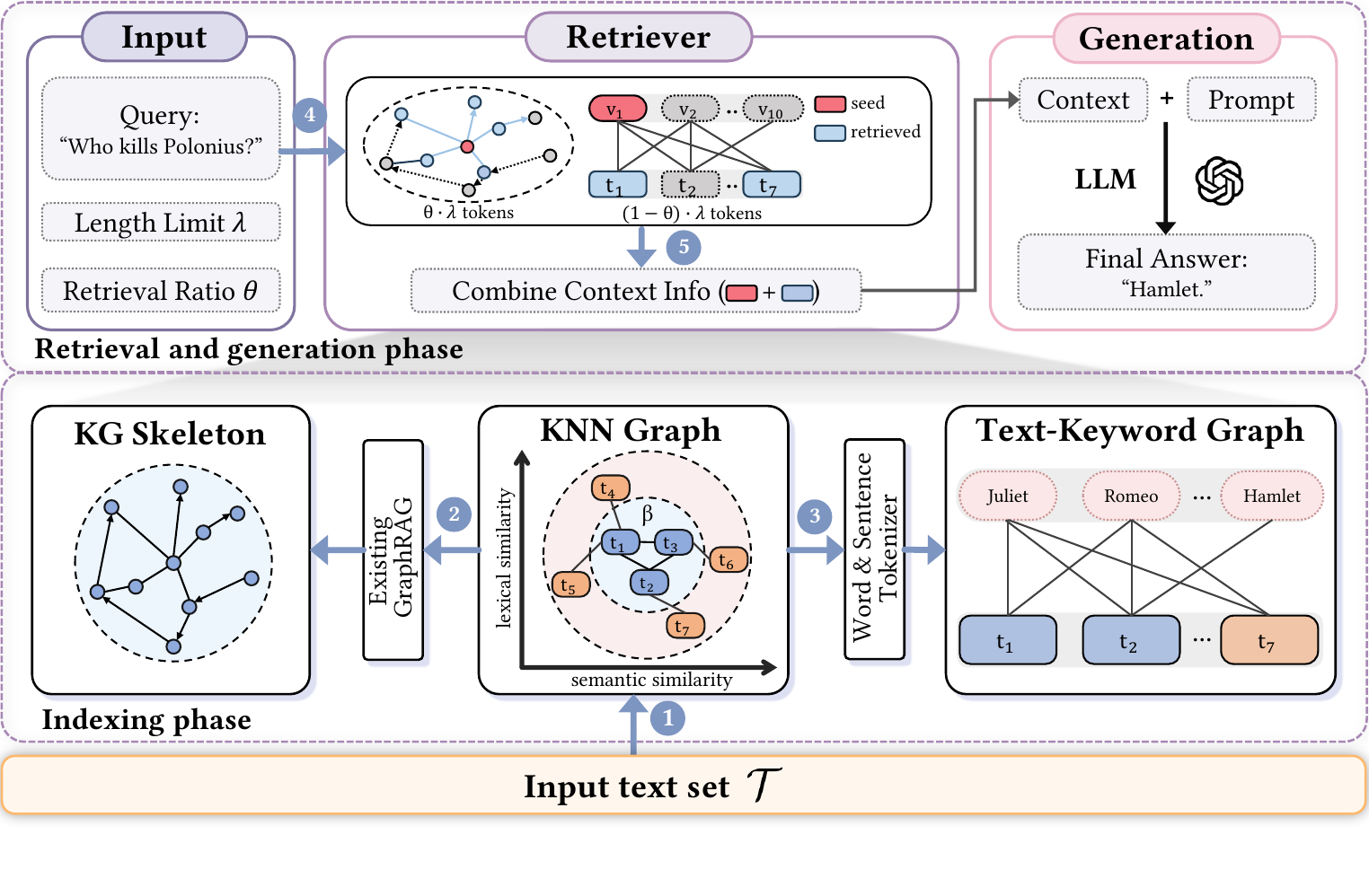}
    \vspace{-9mm}
    \caption{The illustration of \sketrag: the indexing stage in \textcircled{1}-\textcircled{3} and the retrieval stage in \textcircled{4}-\textcircled{5}.}
    \label{fig:sketrag-overview}
    \vspace{-2mm}
\end{figure*}

\section{Proposed Framework: \sketrag}
To fully leverage the strengths of existing graph indices while addressing their limitations, we introduce \sketrag, an indexing framework that integrates multiple levels of granularity: \underline{K}eywords, \underline{E}ntities, and \underline{T}ext chunks, into the \textgraph{} index $\G$. The overall workflow of \sketrag is illustrated in Figure~\ref{fig:sketrag-overview}.

\subsection{Overview}
At a high level, the \textgraph{} index $\G = \G_s \cup \G_k$ consists of: (i) a knowledge graph \textit{skeleton} $\G_s$, derived from a selected set of important text chunks called \textit{core chunks}, and (ii) a text-keyword bipartite graph $\G_k$, constructed from all chunks. As shown in Figure~\ref{fig:sketrag-overview}, the construction process involves three main steps.
\begin{enumerate}
[topsep=2pt,itemsep=1pt,parsep=0pt,partopsep=0pt,leftmargin=15pt]
    \item \sketrag first organizes the input text chunks in $\Tset$ into a KNN graph, where chunks are linked if they exhibit sufficient lexical or semantic similarity. This serves as an intermediate structure for building the final graph $\G$. 
    \item Next, \sketrag selects a $\beta$ fraction of \textit{core chunks} according to their structural importance in the KNN graph. These core chunks are then processed using \graphindex (Algorithm~\ref{alg:graphrag-index}) to produce a knowledge graph skeleton $\G_s$. 
    \item Finally, \sketrag constructs the bipartite graph $\G_k = (\V_k \cup \V_t, \E_k)$ from $\Tset$. In $\G_k$, the node set $\V_k$ represents keywords, and $\V_t$ represents text chunks. An edge $e_{i,j} \in \E_k$ indicates that keyword node $v_i$ appears in text chunk node $v_j$. Each keyword node $v_i$ is assigned a description $t_i$ (consisting of all sentences containing that keyword), and its embedding $\temb{t_i}$ is computed as the average of these sentences' embeddings.
\end{enumerate}
During retrieval, \sketrag balances information from $\G_s$ and $\G_k$ using a constant $\theta$. It first identifies a set of \textit{seed nodes}, either entities or keywords, that are most similar to the query $q$ in the text embedding space. For entity seeds, \sketrag applies Algorithm~\ref{alg:graphrag-retrieval} to retrieve context using $\theta$ proportion of the total context limit $\lambda$. For keyword seeds, it follows a similar procedure to collect relevant neighboring text chunks using the remaining $(1-\theta)$ of the context budget. Finally, the retrieved context is combined with a predefined prompt and passed to the LLM for response generation.

\input{figures/degree-distribution}

\subsection{Rationale and Comparison}\label{sec:ket-rationale}

\sketrag is motivated by two key observations. First, a small subset of core text chunks often exhibits broad relevance to others. Figure~\ref{fig:degree-distribution} presents the degree distribution of the KNN graph constructed from the MuSiQue dataset with input chunk sizes $\clen = 1200$ and $\clen = 150$. This heavily skewed distribution highlights the importance of core chunks in linking different parts of the graph. Consequently, these core chunks should be prioritized to extract high-quality triplets using the LLM.
Second, in the lightweight alternative graph $\G_k$, keywords and their neighboring text chunks can serve as stand-ins for entities and their ego networks. Specifically, when seed keywords align with seed entities, their neighboring text chunks are expected to contain information about those entities' ego networks. Hence, these neighboring chunks are treated as candidates, and retrieval follows the standard \textrag strategy.

To summarize, compared to previous \kgrag solutions~\cite{li2024dalk,delile2024graph,edge2024local,gutierrez2024hipporag}, \sketrag focuses on a smaller set of core chunks to construct a knowledge graph skeleton while leveraging a text-keyword bipartite graph as a lightweight alternative. This design lowers the cost of LLM inference and improves result quality via two distinct retrieval channels (entity and keyword). Additionally, in the keyword channel, \sketrag confines the retrieval to snippets containing seed keywords, unlike \textrag, which searches across the entire $\Tset$. This subgraph-based approach better captures in-text relationships w.r.t.\ seed keywords, enhancing overall effectiveness.

%% file: figures/degree-distribution.tex
\begin{figure}[!t]
\centering
\begin{tikzpicture}
    \begin{customlegend}[legend columns=3,
        legend entries={$K=2$,$K=4$,$K=10$}
        ,
        legend style={at={(0.45,1.05)},anchor=north,draw=none,font=\footnotesize,column sep=0.1cm}]
    \addlegendimage{only marks, line width=0.23mm,mark size=2.3pt,mark=x,color=LightBlue}
    \addlegendimage{only marks, line width=0.23mm,mark size=2.3pt,mark=star,color=Orange}
    \addlegendimage{only marks, line width=0.23mm,mark size=2.3pt,mark=pentagon,color=Green}
    \end{customlegend}
\end{tikzpicture}
\\[-\lineskip]

\subfloat[$\clen=1200$]{
\begin{tikzpicture}[scale=1]
    \begin{axis}[
        height=\columnwidth/2.4,
        width=\columnwidth/1.9,
        ylabel={\#chunks},
        xlabel={degree},
        xmode=log, ymode=log,
        xmin=1, xmax=100,
        ymin=1, ymax=316,
        xtick={1, 10, 100},
        xticklabels={1, 10, 100},
        ytick={1, 10, 100},
        yticklabels={$10^0$, $10^1$, $10^2$},
        every axis y label/.style={at={(current axis.north west)},right=4mm,above=0mm},
        label style={font=\footnotesize},
        tick label style={font=\footnotesize},
        every axis x label/.style={at={(current axis.south)},right=0mm,above=-7mm},
        label style={font=\footnotesize},
        tick label style={font=\footnotesize},
    ]

    \addplot[only marks, mark size=2.3pt, mark=x,color=LightBlue, line width=0.23mm]
        plot coordinates { 
(1, 24)
(2, 275)
(3, 213)
(4, 91)
(5, 45)
(6, 19)
(7, 7)
(8, 2)
(9, 2)
(10, 3)
(14, 1)
    };

    \addplot[only marks, mark size=2.3pt, mark=star,color=Orange, line width=0.23mm]
        plot coordinates { 
(3, 48)
(4, 231)
(5, 157)
(6, 84)
(7, 53)
(8, 30)
(9, 32)
(10, 14)
(11, 10)
(12, 6)
(13, 5)
(14, 5)
(15, 4)
(17, 1)
(22, 1)
(27, 1)
(30, 1)
(31, 1)
    };

    \addplot[only marks, mark size=2.3pt, mark=pentagon,color=Green, line width=0.23mm]
        plot coordinates { 
(7, 1)
(8, 15)
(9, 60)
(10, 128)
(11, 94)
(12, 80)
(13, 46)
(14, 44)
(15, 36)
(16, 12)
(17, 21)
(18, 19)
(19, 15)
(20, 14)
(21, 15)
(22, 8)
(23, 15)
(24, 7)
(25, 5)
(26, 5)
(27, 3)
(28, 6)
(30, 1)
(31, 6)
(32, 6)
(33, 3)
(34, 2)
(35, 2)
(36, 1)
(37, 2)
(38, 1)
(40, 1)
(41, 1)
(42, 2)
(46, 1)
(47, 1)
(50, 1)
(56, 1)
(63, 1)
(67, 1)
(76, 1)
    };

    \end{axis}

\end{tikzpicture}
}%
\hspace{2mm}
\subfloat[$\clen=150$]{
\begin{tikzpicture}[scale=1]
    \begin{axis}[
        height=\columnwidth/2.4,
        width=\columnwidth/1.9,
        ylabel={\#chunks},
        xlabel={degree},
        xmode=log, ymode=log,
        xmin=1, xmax=100,
        ymin=1, ymax=2000,
        xtick={1, 10, 100},
        xticklabels={1, 10, 100},
        ytick={1, 10, 100, 1000},
        yticklabels={$10^0$, $10^1$, $10^2$, $10^3$},
        every axis y label/.style={at={(current axis.north west)},right=4mm,above=0mm},
        label style={font=\footnotesize},
        tick label style={font=\footnotesize},
        every axis x label/.style={at={(current axis.south)},right=0mm,above=-7mm},
        label style={font=\footnotesize},
        tick label style={font=\footnotesize},
    ]

    \addplot[only marks, mark size=2.3pt, mark=x,color=LightBlue, line width=0.23mm]
        plot coordinates { 
(1, 823)
(2, 2427)
(3, 1230)
(4, 547)
(5, 227)
(6, 116)
(7, 57)
(8, 17)
(9, 10)
(10, 5)
(11, 3)
(12, 1)
(13, 1)
(14, 2)
(15, 1)
(20, 1)
    };

    \addplot[only marks, mark size=2.3pt, mark=star,color=Orange, line width=0.23mm]
        plot coordinates { 
(2, 81)
(3, 711)
(4, 1628)
(5, 1138)
(6, 692)
(7, 469)
(8, 261)
(9, 156)
(10, 104)
(11, 81)
(12, 40)
(13, 39)
(14, 18)
(15, 13)
(16, 7)
(17, 10)
(18, 5)
(19, 4)
(20, 1)
(21, 1)
(23, 2)
(24, 1)
(25, 1)
(27, 2)
(29, 1)
(30, 1)
(44, 1)
    };

    \addplot[only marks, mark size=2.3pt, mark=pentagon,color=Green, line width=0.23mm]
        plot coordinates { 
(6, 4)
(7, 44)
(8, 185)
(9, 494)
(10, 907)
(11, 722)
(12, 601)
(13, 495)
(14, 375)
(15, 308)
(16, 263)
(17, 194)
(18, 141)
(19, 132)
(20, 103)
(21, 91)
(22, 71)
(23, 58)
(24, 39)
(25, 33)
(26, 36)
(27, 25)
(28, 20)
(29, 15)
(30, 8)
(31, 14)
(32, 9)
(33, 13)
(34, 10)
(35, 10)
(36, 8)
(37, 4)
(38, 5)
(39, 2)
(40, 2)
(41, 3)
(42, 3)
(43, 2)
(44, 1)
(45, 3)
(46, 2)
(48, 1)
(51, 1)
(52, 2)
(53, 2)
(54, 1)
(55, 1)
(63, 1)
(65, 1)
(69, 1)
(74, 1)
(99, 1)
    };

    \end{axis}

\end{tikzpicture}
}%
\caption{Log-log Plot of the degree distribution of the KNN graph on MuSiQue.}
\label{fig:degree-distribution}
\vspace{-4mm}
\end{figure}
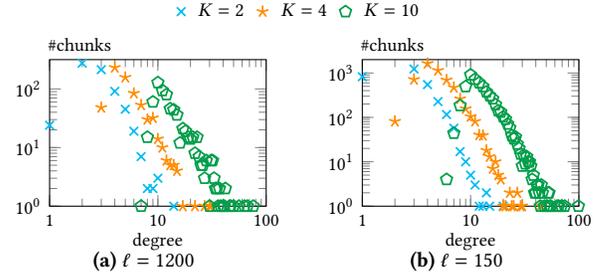

%% file: sections/details.tex
\section{Detailed Implementations}\label{sec:details}
This section provides a detailed explanation of \sketrag, with the indexing stage \sketindex discussed in Section~\ref{sec:graph-construct} and the retrieval process \sketretrieval described in Section~\ref{sec:subgraph-retrieval}.

\begin{algorithm}[!t]
\KwIn{The text chunk set $\Tset$, an integer $K$, a budget rate $\beta$, the number of splits $\tau$.}
\KwOut{A \textgraph{} index $\G$.}

$\W \gets$ all keywords tokenized from $\Tset$\;

$\V \gets \{v_i = (t_i, \temb{t_i}) ~|~ t_i \in \Tset\}; ~ \E \gets \emptyset$\;
\For{each $v_i \in \V$}{
    $\Sset_1 \gets \argmaxk{v_j\in \V \setminus \{v_i\}} \text{co-occ}(v_i,v_j)$, where $k=K/2$\;
    $\Sset_2 \gets \argmaxk{v_j\in \V \setminus (\Sset_1 \cup \{v_i\})} \text{cos}(v_i,v_j)$, where $k=K/2$\;
    $\E \gets \E \cup \{(v_i, v_j) ~|~ v_j \in (\Sset_1 \cup \Sset_2)\}$;
}
$\G \gets (\V, \E)$\;

$\V_c \gets \argmaxk{v_i\in \V} \pi_i$ in Eq.~\eqref{eq:pr}, where $k=\lceil \beta \cdot |\V| \rceil$\;

$\G_s \gets $ \graphindex ($\V_c$) in Algorithm~\ref{alg:graphrag-index}\;
$\Tset_{\tau} \gets$ split each $t_i \in \Tset$ into $2^{\tau}$ equal-sized sub-chunks\;
$\V_t \gets \{v_i = (t_i, \temb{t_i}) ~|~ t_i \in \Tset_{\tau}\}; ~ \V_k \gets \emptyset; ~ \E_k \gets \emptyset$\;

\For{each keyword $x \in \W$}{
    $\V_k \gets \V_k \cup \{v_j = (t_j, \temb{t_j})\}$, where $t_j$ is all sentences in $\Tset_{\tau}$ containing $x$ and $\temb{t_j}$ is the average embedding\;
    $\E_k \gets \E_k \cup \{(v_i, v_j) \mid v_i \in \V_t,\ v_j \in \V_k,\ t_i \text{ contains } x\}$\;
}
$\G_k \gets (\V_k \cup \V_t, \E_k)$\;
\Return{$\G_s\cup \G_k$;}
\caption{\sketindex$(\Tset, K, \beta, \tau)$}
\label{alg:sketrag-index}
\end{algorithm}

\subsection{\sketindex}\label{sec:graph-construct}

As outlined in Algorithm~\ref{alg:sketrag-index}, \sketindex takes as input a set $\Tset$ of text chunks, an integer $K$ for the KNN graph, a budget rate $\beta$, and an integer $\tau$ for text chunk splitting. It first tokenizes all text chunks in $\Tset$ into vocabulary $\W$. By default, it tokenizes chunks into words while excluding stop words (\eg `the', `a', and `is') to define keyword nodes, though traditional keyword extraction methods~\cite{ramos2003using} can also be applied. \sketindex then executes three core subroutines.

\stitle{KNN Graph Initialization}
In Lines 2–7, \sketindex represents each text chunk $t_i \in \Tset$ with its embedding $\temb{t_i}$ as a node $v_i \in \V$. It then links each node $v_i$ to the top-$K/2$ nodes based on lexical similarity and the top-$K/2$ nodes based on semantic similarity, forming the KNN graph $\G$ (Lines 3–6). Specifically, the lexical similarity between nodes $v_i, v_j$ is defined as the number of co-occurring keywords in $\W$, while the semantic similarity is measured using the cosine similarity between their embeddings $\temb{t_i}$ and $\temb{t_j}$.

\stitle{core chunk identification}
Motivated by the observations in Figure~\ref{fig:degree-distribution}, given an intermediate KNN graph $\G = (\V, \E)$ and a budget rate $\beta \in [0,1]$, Line 8 of Algorithm~\ref{alg:sketrag-index} selects a set $\V_c$ of $\lceil \beta\cdot|\V|\rceil$ core chunk nodes based on their structural importance.
For each node $v_i \in \V$, the PageRank value~\cite{page1999pagerank} $\pi_i$ serves as a measure of structural importance. Let $\mathbf{P}$ be the probability transition matrix of $\G$, where $\mathbf{P}_{i,j} = \frac{1}{d(v_i)}$ and $d(v_i)$ denotes the degree of $v_i$. Given a teleport probability $\alpha$, the PageRank vector $\boldsymbol{\pi}$ is computed as:
\begin{equation}\label{eq:pr}
\boldsymbol{\pi} = \alpha\cdot\mathbf{1/n} + (1-\alpha)\cdot \boldsymbol{\pi} \cdot\mathbf{P},
\end{equation}  
where $\mathbf{1/n}$ is the initial vector with each of $n = |\V|$ dimensions set to $1/n$, and $\pi_i = \boldsymbol{\pi}[i]$ represents the PageRank score of $v_i \in \V$. PageRank effectively captures both direct and higher-order structural importance, making it suitable for identifying core chunks.

\stitle{Graph Index Construction}
In Line 9, \sketindex processes the selected core text chunks $\V_c$ using \graphindex (Algorithm~\ref{alg:graphrag-index}) to construct the knowledge graph skeleton $\G_s$. Since the text-attributed node set $\V_c$ has already been built, Line 1 in Algorithm~\ref{alg:graphrag-index} is skipped.
Next, in Lines 10--15, \sketindex constructs a text-keyword bipartite graph $\G_k = (\V_k \cup \V_t, \E_k)$ based on $\Tset$. Specifically, each text chunk in $\Tset$ is recursively divided into equal-sized sub-chunks over $\tau$ iterations, forming a set $\Tset_{\tau}$ with $2^{\tau}\cdot|\Tset|$ sub-chunks. Each sub-chunk is then initialized to a node in $\V_t$. 
For each keyword $x\in \W$, \sketindex creates a keyword node $v_{j}=(t_j,\temb{t_j})$, where $t_j$ concatenates all sentences in $\Tset$ containing $x$, and $\temb{t_j}$ is their average embedding. This process aggregates information from different contexts, reducing noise and generating a more generalized representation of the keyword. Finally, \sketindex links each chunk node $v_i \in \V_t$ to its corresponding keyword node $v_j$.
After constructing $\G_k$, \sketindex returns $\G_s \cup \G_k$ as the final index $\G$, where text chunk nodes in $\G_s$ are replaced by their corresponding partitioned sub-chunk nodes in $\V_t$ of $\G_k$, with edges in $\G_s$ rewired accordingly.

\subsection{\sketretrieval}\label{sec:subgraph-retrieval}
As outlined in Algorithm~\ref{alg:sketrag-retrieval}, \sketretrieval takes as input a \textgraph{} index $\G = \G_s \cup \G_k$, a query $q$, a context length limit $\lambda$, and a retrieval ratio $\theta$. It outputs a context $C$ by selecting the most relevant content from the skeleton graph $\G_s$ and the keyword-text bipartite graph $\G_k$.
In Line 1, \sketretrieval invokes \graphretrieval (Algorithm~\ref{alg:graphrag-retrieval}) to retrieve context $C_s$ from $\G_s$, using $\theta \cdot \lambda$ tokens. Next, in Lines 2–4, it retrieves content from $\G_k$ using the remaining $(1-\theta) \cdot \lambda$ tokens.
Specifically, \sketretrieval iteratively selects seed keyword nodes $\Sset_k$ from $\V_k$ based on cosine similarity to the query embedding, expanding the selection until their neighboring text sub-chunks contain $2 \cdot (1-\theta) \cdot \lambda$ tokens, \ie $\left|\bigoplus\left(\N\left(\Sset_k\right)\right)\right| = 2\cdot(1-\theta)\cdot\lambda$.
Focusing on the candidate set $\N\left(\Sset_k\right)$, \sketretrieval retrieves the final text set $\Sset_t$ based on the cosine similarity. These texts are concatenated into context $C_k$ with $(1-\theta) \cdot \lambda$ tokens. Finally, it returns $C_s\oplus C_k$ as the retrieved context for \sketrag.

Notably, all chunks retrieved by \sketretrieval, whether from entity or keyword channels, are fine-grained sub-chunks generated during the indexing stage through spitting and rewiring. This refinement reduces noise and preserves the context limit, allowing for the retrieval of more relevant knowledge during online queries.

\begin{algorithm}[!t]
\KwIn{A \textgraph{} index $\G = \G_s \cup \G_k$, a query $q$, context length limit $\lambda$, retrieval ratio $\theta$}
\KwOut{The context $C$}
$C_{s} \gets \graphretrieval \left(\G_s, \theta \cdot \lambda\right)$ in Algorithm~\ref{alg:graphrag-retrieval}\;

$\Sset_k \gets \argmaxk{v_{i}\in \V_k} \text{cos}(v_i,q)$, s.t.\ ${\scriptstyle \left|\bigoplus\left(\N\left(\Sset_k\right)\right)\right| = 2\cdot(1-\theta)\cdot\lambda}$\;
$\Sset_t \gets \argmaxk{v_{i}\in \N(\Sset_k)} \text{cos}(v_i,q)$, s.t.\ ${\scriptstyle \left|\bigoplus(\Sset)\right| = (1-\theta)\cdot\lambda}$\;
$C_k\gets \bigoplus(\Sset)$\;
\Return{$C_s \oplus C_k$;} 
\caption{\sketretrieval$(\G, q, \lambda, \theta)$}
\label{alg:sketrag-retrieval}
\end{algorithm}

\subsection{Cost Analysis}\label{sec:analysis}

We begin by analyzing the cost of \graphindex to construct $\G = (\V_e\cup\V_t, \E)$. Let $\lambda_e$ and $\lambda_r$ denote the token counts of the prompt templates used to extract entities and relationships, respectively. Each text chunk node $v_i \in \V_t$ has a text of length $\clen$. To extract entities, the LLM is prompted with $\clen + \lambda_e$ tokens for each $v_i$, resulting in $(\clen + \lambda_e)\cdot|\V_t|$ total input tokens. Similarly, extracting relationships requires $(\clen + \lambda_r)\cdot|\V_t|$ tokens. In addition, \graphindex must compute text embeddings for all nodes and edges, incurring another $\clen\cdot|\V_t|+\sum\limits_{x\in \V_e\cup\E}\ell_x$ tokens, where $\ell_x$ is the description length of $x\in \V_e\cup\E$. Therefore, the total LLM Input Token Cost (ITC) for \graphindex is:
$$\text{ITC}_{\graphindex} = \left(2+\frac{(\lambda_e+\lambda_r)}{\clen}\right)\cdot\clen\cdot|\V_t|\cdot c_i + \left(\clen\cdot|\V_t|+\sum\limits_{x\in \V_e\cup\E}\ell_x\right)\cdot c_e,$$
where $c_i$ and $c_e$ are the per-token costs for the LLM and embedding models, respectively.
By contrast, \sketindex uses only a $\beta$-fraction of the \graphindex input token cost to construct $\G_s$. To build the text-keyword bipartite graph $\G_k$, it additionally consumes $3\clen\cdot|\Tset|$ tokens for multi-granular text embeddings (chunk, sub-chunk, and sentence levels). Thus,
$$\text{ITC}_{\sketindex} = \beta\cdot \text{ITC}_{\graphindex} + 3\cdot\clen\cdot|\Tset|\cdot c_e.$$
Regarding output token costs, \sketindex incurs only a $\beta$ fraction of the output cost of \graphindex, as it generates only $\beta$ of the entities and relations present in the full knowledge graph.

Regarding the retrieval and generation stages, all solutions, including \sketrag, incur the same upper-bounded cost, as both stages are regulated by the maximum token parameter during LLM inference. Specifically, the input tokens for all solutions comprise a distinct prompt template and the retrieved content, which is constrained by the limit $\lambda$. The number of output tokens is then determined by subtracting the input token count from the maximum token limit, ensuring consistent computational costs across different approaches.

%% file: sections/exp.tex
\section{Experiments}\label{ref:exp}
In this section, we evaluate our proposed \sketrag framework by addressing the following research questions:
\begin{itemize}[topsep=2pt,itemsep=1pt,parsep=0pt,partopsep=0pt,leftmargin=11pt]
\item \textbf{RQ1}: How does \sketrag enhance effectiveness while reducing costs compared to existing solutions?
\item \textbf{RQ2}: What is the contribution of each core subroutine to \sketrag's overall performance?
\item \textbf{RQ3}: How does \sketrag balance result quality, index construction cost, and the trade-off between its two retrieval channels?
\item \textbf{RQ4}: How sensitive is \sketrag's performance to its parameter settings?
\item \revise{\textbf{RQ5}: How does \sketrag’s retrieval latency compare to other baselines, and is there any compromise?}
\end{itemize}
All experiments are conducted on a Linux machine with Intel Xeon(R) Gold 6240@2.60GHz CPU and 32GB RAM. We use the OpenAI API to access LLMs.

\input{figures/overall-table-low}

\subsection{Experimental Settings}\label{sec:exp-set}

\stitle{Datasets and metrics}
We use two widely adopted benchmarking datasets for multi-hop QA tasks, MuSiQue~\cite{trivedi2022musique} and HotpotQA~\cite{yang2018hotpotqa}, \revise{together with a benchmark RAG-QA Arena~\cite{han-etal-2024-rag} for open-ended RAG evaluation.} These datasets consist of QA pairs, each accompanied by multiple paragraphs as potential relevant context. Specifically, each QA pair is associated with 2 golden paragraphs out of 20 candidate paragraphs in MuSiQue, 2 golden paragraphs and 8 distracting paragraphs in HotpotQA, \revise{and one or more relevant paragraphs in RAG-QA Arena}. Following prior work~\cite{wang2024knowledge, gutierrez2024hipporag}, we sample 500 QA instances from each dataset. The corresponding paragraphs for all sampled instances are preprocessed into $\Tset$ and compiled as the external corpus for RAG~\cite{gutierrez2024hipporag}.
\revise{The number of preprocessed paragraphs is 6,761 for MuSiQue (resp.\ 20,150 for HotpotQA and 7,010 for RAG-QA Arena), resulting in an overall token count of 751,784 (resp.\ 618,325 and 1,357,333).}
\revise{For MuSiQue and HotpotQA, we assess retrieval quality using \textit{Coverage}, defined as the proportion of QA instances where the ground-truth answer is found within the retrieved context. Generation quality is evaluated by generating answers using a pre-defined LLM based {\it solely} on the retrieved context, and comparing these answers against ground truths using the following standard metrics~\cite{yu2024evaluation, gutierrez2024hipporag,wang2024knowledge,li2024graph}: \textit{Exact Match (EM)}, which measures the percentage of exact matches with the ground truth; \textit{F1 score}, which captures partial correctness via word-level overlap; and \textit{BERTScore}, which computes semantic similarity using BERT-based embeddings. For RAG-QA Arena, where answers are long-form, generation quality is evaluated using the Win Rate metric~\cite{han-etal-2024-rag}, defined as the proportion of cases in which the generated answer is preferred over a baseline answer (\ie \textrag with $\ell=1,200$) by an LLM. We use OpenAI's \texttt{GPT-4o-mini} as the judgment LLM.
}

\stitle{Solutions and configurations}
\revise{We evaluate the performance of 13 solutions categorized as follows:} (i) \textit{Existing competitors}: \textrag~\cite{lewis2020retrieval}, \knnrag~\cite{wang2024knowledge}, \msgraphrag~\cite{edge2024local}, \hybridrag~\cite{sarmah2024hybridrag}, \revise{\hyde~\cite{gao2023precise}, \hipporag~\cite{gutierrez2024hipporag}, \lightrag~\cite{guo2025lightrag} including three retrieval frameworks \lightragl, \lightragg, and \lightragh}. (ii) \textit{Proposed baselines}: \keyrag, which constructs only $\G_k$ as the index and retrieves from it; \skeletonrag, which retains only $\G_s$. (iii) \textit{Final solutions}: \sketragu and \sketragp, where \textsf{U} and \textsf{P} represent selecting core chunks randomly and via PageRank, respectively. For a fair comparison, we implement \revise{most} existing solutions within the \sketrag framework, demonstrating that our approach serves as a unified and generalized RAG framework. Specifically, for \textrag and \knnrag, we use the KNN graph constructed in Algorithm~\ref{alg:sketrag-index} as the index $\G$. \sketrag reduces to \textrag when retrieving only the seed nodes in $\G$ and to \knnrag when also including their neighbors. Furthermore, \sketrag simplifies to \msgraphrag by setting $\beta=1$ and $\theta=1$ and to \hybridrag by combining both \textrag and \msgraphrag. Notably, \hybridrag fully constructs indices for both \textrag and \msgraphrag, retrieving content equally from both. This setup effectively corresponds to $\beta=0.5$ under a fixed context limit. \revise{For \hyde, we implement it by altering the generation prompt based on \textrag. Notably, for \lightrag and \hipporag, we use their original code while aligning the same input parameters in subsequent experiments.}
Regarding the base models in each solution, we use OpenAI's \texttt{GPT-4o-mini} as the LLM for inference, OpenAI's \texttt{text-embedding-3-small} for text embedding generation, \texttt{cl100k\_base} for word tokenization, and the \texttt{sent\_tokenize} function from the \texttt{nltk} Python library for sentence tokenization.
We follow the default settings in~\citet{edge2024local}, setting the input chunk size $\ell$ to 1,200 and the output context limit $\lambda$ to 12,000 across all solutions. Within \sketrag, we use the default parameters for components related to \kgrag and set $K=2$, $\beta=0.8$, and $\theta=0.4$, unless specified otherwise.
The implementations of all solutions are available at {\color{blue}\url{https://github.com/waetr/KET-RAG}}.

\input{figures/overall-table-high}

\subsection{Performance Evaluation (RQ1)}
In the first set of experiments, we evaluate the performance of \sketrag against existing competitors \eat{(\textrag, \knnrag, \msgraphrag, and \hybridrag)} under two configurations: a low-cost version with reduced accuracy and a high-accuracy version with increased cost. Following previous works~\cite{edge2024local}, we achieve the low-cost setting by using an input chunk size of $\ell = 1,200$ and the high-accuracy setting by fixing $\ell = 150$ for all solutions. For \keyrag and \sketrag, we set $\tau$ to 3 and 0, respectively, ensuring the same text sub-chunk length in both configurations.

As reported in Tables~\ref{tab:quality-all-low}-\ref{tab:quality-all-high}, our proposed \sketrag(\textsf{-U}/\textsf{-P}) achieves superior quality-cost trade-offs compared to existing methods across all datasets.
In terms of retrieval quality, \sketrag~\revise {ranks as the best or second-best}, achieving coverage scores of 77.0\%/80.2\% and 81.6\%/82.6\% on MuSiQue and HotpotQA, respectively. Compared to the competitor \revise{\lightrag (including \lightragg, \lightragl, and \lightragh)}, this corresponds to relative improvements of 68.8\% and 32.5\% on MuSiQue and HotpotQA, respectively. Most notably, we observe that \sketrag in low-cost mode achieves comparable or even superior coverage to \msgraphrag, \revise{\hipporag, and \lightrag} in high-accuracy mode while reducing indexing costs by over an order of magnitude. For example, on HotpotQA, the coverage scores of \sketragp, \msgraphrag, \revise{\hipporag, and \lightrag} are 81.6\%, 74.6\%, 79.6\%, and 76.8\%, respectively, yet \sketragp incurs only 18.3\% of their indexing cost.
Akin to the retrieval quality, \sketrag achieves competitive generation quality at lower costs. Take the low-cost mode as an example. It improves \hybridrag by 37.2\%/31.3\%/0.3\% (resp.\ 26.5\%/26.9\%/0.9\%) in EM/F1 score/BERTScore on MuSiQue (resp.\ HotpotQA) while reducing indexing costs by $19\%$; \revise{on RAG-QA Arena, it improves the best competitor \hipporag by 3.1\%.}

Regarding other competitors, we observe that \textrag exhibits a more pronounced accuracy improvement compared to \msgraphrag, \revise{\hipporag, and \lightrag} when transitioning from low- to high-accuracy mode, which motivates the text splitting strategy in \sketrag.
Additionally, we find that, on HotpotQA, the generation quality of \msgraphrag is slightly lower than that of \textrag in the high-accuracy setting. This is because HotpotQA is a weaker benchmark for multi-hop reasoning due to the presence of spurious signals~\cite{gutierrez2024hipporag, trivedi2022musique}. Despite this, \sketrag consistently outperforms both competitors, demonstrating its robustness across different knowledge retrieval scenarios.

\input{figures/fig-beta}
\input{figures/fig-theta}

\subsection{Ablation Study (RQ2)}
In the second set of experiments, we evaluate the performance of each single building block proposed in \sketrag, whose results are also included in Tables~\ref{tab:quality-all-low}-\ref{tab:quality-all-high}.

\stitle{Knowledge graph skeleton}
We evaluate the performance of \skeletonrag with its full version, \msgraphrag. By default, \skeletonrag sets $\beta=0.8$, resulting in a 20\% reduction in indexing cost across all cases. 
Surprisingly, we find that \skeletonrag trades off only minor performance reductions, particularly in low-cost settings, while maintaining parity in high-accuracy configurations. For instance, in terms of EM score, \skeletonrag exhibits a relative decrease of 3.5\% and 7.4\% in the low-cost setting on MuSiQue and HotpotQA, respectively. However, in high-accuracy settings, there is no performance drop, and in the case of HotpotQA, even a slight improvement is observed. These results suggest that \skeletonrag effectively balances efficiency and quality, making it a viable alternative to full-scale knowledge graph indexing.

\stitle{Text-keyword bipartite graph}
We compare \keyrag with the conventional \textrag. To ensure a fair comparison, we set $\tau=0$ for \keyrag, allowing both \keyrag and \textrag to retrieve text chunks of the same size. As shown in Tables~\ref{tab:quality-all-low}-\ref{tab:quality-all-high}, \keyrag consistently outperforms \textrag in retrieval and generation quality. Notably, in the low-cost setting, \keyrag achieves 92.4\%/133.3\%/118.5\% and 61.8\%/52.5\%/58.8\% relative improvement in Coverage/EM/F1 on MuSiQue and HotpotQA, respectively, with more significant gains on MuSiQue. It demonstrates the effectiveness of retrieving context from neighboring text chunks of keyword seeds, particularly in complex multi-hop reasoning.

\stitle{Core text chunk identification}
Based on the results in Tables~\ref{tab:quality-all-low}-\ref{tab:quality-all-high}, we observe that \sketragp shows better quality than \sketragu in both the low-cost and high-cost modes. For instance, \sketragp outperforms \sketragu by up to 1.0\%/4.5\%/4.4\% in Coverage/EM/F1 scores on MuSiQue.
This confirms the effectiveness of the core chunk identification technique, as motivated in Section~\ref{sec:ket-rationale}.
Additionally, we observe that the superiority of \sketragp remains consistent across different settings of $\beta$ and $\theta$, as further illustrated in Figures~\ref{fig:quality-beta}–\ref{fig:quality-theta}.

\subsection{Trade-off Analysis (RQ3)}
In the third set of experiments, we analyze the trade-off between accuracy and cost by varying the budget $\beta$, as well as the balance between the two retrieval channels by adjusting $\theta$. We set $\clen = 150$ and $\tau=0$, and follow the default parameter settings in Section~\ref{sec:exp-set}.

\stitle{Accuracy-cost tradeoff}
Figure~\ref{fig:quality-beta} presents the performance of \sketragu, \sketragp, and \skeletonrag on MuSiQue and HotpotQA by varying $\beta$. 
In particular, \sketragp and \sketragu consistently outperform \skeletonrag, which serves as a lower bound, across different budget $\beta$ values. 
Between the two variants, \sketragp achieves better performance than \sketragu particularly when $\beta\in [0.6, 0.8]$ in MuSiQue and $\beta\in [0.2, 0.4]$ in MuSiQue, demonstrating the effectiveness of identifying core text chunks using PageRank centralities.
Furthermore, the performance of \sketrag is less sensitive to variations in $\beta$ on HotPotQA. For instance, the coverage at $\beta=0.2$ decreases by 2\% compared to $\beta=1$. On MuSiQue, although \sketrag exhibits greater sensitivity in generation quality, its performance quickly catches up once $\beta$ reaches $0.6$. These findings demonstrate \sketrag's effectiveness for further reducing indexing costs.

\stitle{Retrieval channel}
Figure~\ref{fig:quality-theta} reports the performance of \sketragu, \sketragp, and \hybridrag by varying $\theta$.
As illustrated, \sketrag consistently outperforms \hybridrag across different $\theta$ settings, demonstrating the effectiveness of its two key components, \keyrag and \skeletonrag.
Additionally, we observe that the performance of \keyrag improves significantly when incorporating a small fraction (\eg 0.2) of context from \msgraphrag. This finding further motivates the reduction of costs associated with building a full knowledge graph.
Regarding the two variants, \sketragu and \sketragp, we find that \sketragp achieves superior EM and F1 scores when $\theta \leq 0.4$, which aligns with the trend observed in Figure~\ref{fig:quality-beta}.

\begin{table}[t]
\centering
\renewcommand{\arraystretch}{1.1}
\begin{small}
\caption{Answer quality by varying $\clen$ and $\tau$ on MuSiQue.}
\vspace{-2mm}
\label{tab:quality-clen}
\begin{tabular}{lcccc|cccc}	
    \toprule
    \bf Param & \multicolumn{4}{c}{\bf $\clen$} & \multicolumn{4}{c}{\bf $\tau$} \\
    \cmidrule(lr){2-5} \cmidrule(lr){6-9}
    \bf Value & \bf 150 & \bf 300 & \bf 600 & \bf 1200 & \bf 3 & \bf 2 & \bf 1 & \bf 0 \\
    \midrule
    \bf Coverage & 79.6 & 79.6 & 77.8 & 77.0 & 77.0 & 70.4 & 61.0 & 56.8 \\
    \bf EM & 19.2 & 18.8 & 15.4 & 14.0 & 14.0 & 13.8 & 11.8 & 12.8\\
    \bf F1 & 22.3 & 18.8 & 17.7 & 17.2 & 18.9 & 18.3 & 16.3 & 17.2\\
    \bottomrule
\end{tabular}
\end{small}
\end{table}

\begin{table}[t]
\centering
\renewcommand{\arraystretch}{1.1}
\begin{small}
\caption{Answer quality by varying $K$ on MuSiQue.}
\vspace{-2mm}
\label{tab:quality-k}
\begin{tabular}{lccc}	
    \toprule
    \bf $K$ & \bf 2 & \bf 4 & \bf 10\\
    \midrule
    \bf Coverage/EM/F1 & 79.6/19.2/26.1 & 80.0/17.8/25.2 & 80.4/19.4/26.0\\
    \bottomrule
\end{tabular}
\end{small}
\end{table}

\subsection{Parameter Sensitivity (RQ4)}

In the fourth set of experiments, we take the Musique dataset as an example and analyze the sensitivity of \sketragp w.r.t.\ the input text chunk size $\clen$, the number of splits $\tau$, and the integer $K$ used for KNN graph construction.
For $\clen$, we vary $\clen = 150, 300, 600, 1200$ and set the corresponding $\tau = 0, 1, 2, 3$ to maintain a consistent sub-chunk length. For $\tau$, we fix $\clen = 1200$ and vary $\tau = 0, 1, 2, 3$. Additionally, we set $K = 2, 4, 10$ to explore different KNN graph densities.
As shown in Table~\ref{tab:quality-clen}, \sketragp achieves better retrieval and generation quality as the size of input chunks or split sub-chunks decreases. This trend is consistent with the performance of \keyrag and \graphrag in both low- and high-cost settings. These findings align with previous observations~\cite{edge2024local} that smaller chunk sizes improve result quality, further validating the text chunk splitting design in \sketrag. 
Additionally, as shown in Table~\ref{tab:quality-k}, varying the integer $K$ from 2 to 10 results in only minor changes in coverage, EM, and F1 scores, \eg 79.6/19.2/26.1 for $K = 2$ vs.\ 80.4/19.4/26.0 for $K = 10$, indicating that \sketrag is not significantly affected by the density of the KNN graph. This stability can be explained by Figure \ref{fig:degree-distribution}, which shows that KNN graphs with different $K$ exhibit similar degree distribution shapes, despite variations in average degree.

\subsection{Retrieval Latency (RQ5)}
In the fifth set of experiments, we report the average retrieval latency per query. To reduce retrieval latency, our framework employs concurrent retrieval using multithreading, with the maximum thread count set to 30. We report the retrieval latency of \sketrag and all other methods implemented within the same unified framework in Table~\ref{tab:retrieve-time}.
Specifically, \sketrag achieves average retrieval times of 0.39$\times$/0.73$\times$ those of \msgraphrag and \hybridrag on MuSiQue, 0.43$\times$/0.80$\times$ on HotpotQA, and 0.41$\times$/0.70$\times$ on RAG-QA Arena. Although \textrag is the fastest due to its simplicity, \sketrag's retrieval latency remains under one second. In the low-cost setting, all graph-based RAG variants complete retrieval within 0.5 seconds. LLM generation time is consistent across methods, averaging 0.42 seconds per query. Overall, there is \textit{no significant compromise} in latency when using \sketrag, especially given its improved retrieval quality.

\begin{table}[!t]
\centering
\renewcommand{\arraystretch}{1.1}
\begin{small}
\caption{Retrieval time (low cost/high perf.) in seconds.}
\vspace{-2mm}
\label{tab:retrieve-time}
\begin{tabular}{l ccc}
\toprule
\bf Method  & \bf MuSiQue & \bf HotpotQA & \bf RAG-QA Arena \\
\midrule
\textrag      & 0.03/0.03   & 0.03/0.02    & 0.03/0.02   \\
\knnrag       & 0.03/0.23   & 0.03/0.18    & 0.04/0.38   \\
\msgraphrag   & 0.43/1.99   & 0.39/1.75    & 0.46/2.14   \\
\hybridrag    & 0.41/1.07   & 0.40/0.93    & 0.43/1.26   \\
\hyde         & 0.04/0.03   & 0.03/0.03    & 0.04/0.02   \\
\keyrag       & 0.03/0.04   & 0.04/0.04    & 0.04/0.05   \\
\skeletonrag  & 0.31/1.64   & 0.28/1.51    & 0.32/1.68   \\
\sketragu     & 0.28/0.79   & 0.29/0.63    & 0.28/0.87   \\
\sketragp     & 0.29/0.78   & 0.40/0.87    & 0.28/0.88   \\
\bottomrule
\end{tabular}
\end{small}
\end{table}

%% file: figures/overall-table-low.tex
\begin{table*}[h]
\centering
\renewcommand{\arraystretch}{1.1}
\begin{small}
\caption{Overall performance of RAG methods in low-cost versions. The best and second‐best results in each column are highlighted in bold and underlined, respectively.}
\label{tab:quality-all-low}
\begin{tabular}{lccccc|ccccc|cc}
\toprule
\bf Dataset 
 & \multicolumn{5}{c}{\bf MuSiQue} 
 & \multicolumn{5}{c}{\bf HotpotQA} 
 & \multicolumn{2}{c}{\bf RAG-QA Arena} \\
\cmidrule(lr){2-6} \cmidrule(lr){7-11} \cmidrule(lr){12-13}
\textbf{Metric} 
 & \bf USD & \bf Coverage & \bf EM & \bf F1 & \bf BERTScore
 & \bf USD & \bf Coverage & \bf EM & \bf F1 & \bf BERTScore
 & \bf USD & \bf Win Rate \\
\midrule
\textrag
 & 0.02 & 26.4 & 3.0  & 5.4  & 62.9
 & 0.01 & 37.2 & 14.8 & 19.4 & 69.9
 & 0.03 & 0.0 \\

\knnrag
 & 0.02 & 21.2 & 2.8  & 4.6  & 62.5
 & 0.01 & 28.6 & 13.4 & 16.9 & 68.8
 & 0.03 & 17.8 \\

\msgraphrag
 & 2.30 & 47.6 & 11.4 & 15.8 & 67.4
 & 2.30 & 63.0 & 21.6 & 30.2 & 74.2
 & 5.05 & 33.4 \\

\hybridrag
 & 2.32 & 49.2 & 10.4 & 15.1 & \underline{68.8}
 & 2.31 & 64.8 & 22.6 & 30.5 & 76.8
 & 5.08 & 35.0 \\

\hyde
 & 0.03 & 33.0 & 4.2  & 6.7  & 63.5
 & 0.02 & 40.8 & 16.2 & 21.3 & 70.9
 & 0.03 & 33.2 \\

\hipporag
 & 1.49 & 51.6 & 9.2  & 14.2 & 66.8
 & 1.40 & 64.0 & \textbf{29.2} & \underline{38.3} & 77.3
 & 1.37 & \underline{38.4} \\

\lightragl
 & 1.80 & 39.0 & 9.4  & 12.9 & 66.6
 & 1.77 & 57.6 & 22.4 & 28.1 & 73.1
 & 4.00 & 29.0 \\

\lightragg
 & 1.80 & 37.6 & 6.4  & 9.5  & 64.8
 & 1.77 & 49.0 & 19.6 & 24.5 & 71.8
 & 4.00 & 23.0 \\

\lightragh
 & 1.80 & 45.6 & 10.2 & 14.4 & 67.2
 & 1.77 & 61.6 & 24.6 & 30.7 & 74.2
 & 4.00 & 30.0 \\
\midrule
\keyrag
 & 0.03 & 50.8 & 7.0  & 11.8 & 68.1
 & 0.03 & 60.2 & 24.4 & 33.5 & \textbf{78.9}
 & 0.07 & 37.0 \\

\skeletonrag
 & 1.86 & 43.4 & 11.0 & 14.1 & 66.8
 & 1.84 & 57.8 & 20.0 & 26.7 & 73.2
 & 4.04 & 28.6 \\

\sketragu
 & 1.89 & \underline{76.2} & \underline{13.4} & \underline{18.1} & 68.1
 & 1.87 & \underline{81.4} & 28.4  & 38.2  & \underline{77.5}
 & 4.08 & {35.0} \\

\sketragp
 & 1.89 & \textbf{77.0} & \textbf{14.0} & \textbf{18.9} & \textbf{69.0}
 & 1.87 & \textbf{81.6} & \underline{28.6} & \textbf{38.7} & \underline{77.5}
 & 4.08 & \textbf{39.6} \\
\bottomrule
\end{tabular}
\end{small}
\end{table*}

%% file: figures/overall-table-high.tex
\begin{table*}[h]
\centering
\renewcommand{\arraystretch}{1.1}
\begin{small}
\caption{Overall performance of RAG methods in high-performance versions. The best and second‐best results in each column are highlighted in bold and underlined, respectively.}
\label{tab:quality-all-high}
\begin{tabular}{lccccc|ccccc|cc}
\toprule
\bf Dataset 
 & \multicolumn{5}{c}{\bf MuSiQue} 
 & \multicolumn{5}{c}{\bf HotpotQA} 
 & \multicolumn{2}{c}{\bf RAG-QA Arena} \\
\cmidrule(lr){2-6} \cmidrule(lr){7-11} \cmidrule(lr){12-13}
\bf Metric 
 & \bf USD & \bf Coverage & \bf EM & \bf F1 & \bf BERTScore
 & \bf USD & \bf Coverage & \bf EM & \bf F1 & \bf BERTScore
 & \bf USD & \bf Win Rate \\
\midrule
\textrag
 & 0.05 & 76.8 & 12.8 & 19.4 & 68.8
 & 0.04 & 74.0 & 33.6 & 44.2 & 79.4
 & 0.09 & 40.8 \\

\knnrag
 & 0.05 & 66.0 & 11.2 & 17.5 & 68.7
 & 0.04 & 63.2 & 29.8 & 39.6 & 78.1
 & 0.09 & 35.6 \\

\msgraphrag
 & 24.94 & 69.6 & 17.4 & 25.1 & 71.0
 & 21.29 & 74.6 & 31.0 & 43.0 & 79.5
 & 46.71 & 35.2 \\

\hybridrag
 & 24.99 & 80.0 & \textbf{19.4} & \underline{26.2} & \underline{71.2}
 & 21.33 & 80.2 & 34.0 & 46.1 & 80.2
 & 46.80 & 39.4 \\

\hyde
 & 0.05 & 76.8 & 15.4 & 22.5 & 70.3
 & 0.05 & 78.8 & \textbf{35.2} & 45.5 & 80.3
 & 0.13 & \underline{43.0} \\

\hipporag
 & 8.70 & \textbf{81.4} & 14.8 & 19.8 & 69.0
 & 10.19 & 79.6 & 31.8 & 43.8 & 79.7
 & 12.21 & 39.6 \\

\lightragl
 & *    & *    & *    & *    & *
 & 11.06 & 76.0 & 31.0 & 43.6 & 79.7
 & 29.24 & 38.4 \\

\lightragg
 & *    & *    & *    & *    & *
 & 11.06 & 63.0 & 20.8 & 30.0 & 74.6
 & 29.24 & 38.2 \\

\lightragh
 & *    & *    & *    & *    & *
 & 11.06 & 76.8 & 29.8 & 40.9 & 78.8
 & 29.24 & 39.8 \\
\midrule
\keyrag
 & 0.09 & 78.2 & 14.6 & 20.6 & 69.3
 & 0.07 & \underline{82.2} & 33.8 & 46.4 & 80.8
 & 0.15 & 41.6 \\

\skeletonrag
 & 19.95 & 69.6 & 17.4 & 24.6 & 70.7
 & 17.03 & 74.6 & 31.2 & 42.8 & 79.3
 & 37.39 & 34.4 \\

\sketragu
 & 20.04 & \underline{80.2} & 18.4 & 25.9 & 71.0
 & 17.10 & \textbf{82.6} & \underline{34.8} & \underline{47.2} & \underline{81.4}
 & 37.53 & 38.0 \\

\sketragp
 & 20.04 & 79.6 & \underline{19.2} & \textbf{26.6} & \textbf{71.3}
 & 17.10 & \textbf{82.6} & \textbf{35.2} & \textbf{47.7} & \textbf{81.5}
 & 37.53 & \textbf{43.2} \\
\bottomrule
\multicolumn{13}{l}{\footnotesize * We exclude the results of \lightrag on MuSiQue due to unexpected behavior observed when running its original implementation.}
\end{tabular}
\end{small}
\end{table*}

%% file: figures/fig-beta.tex
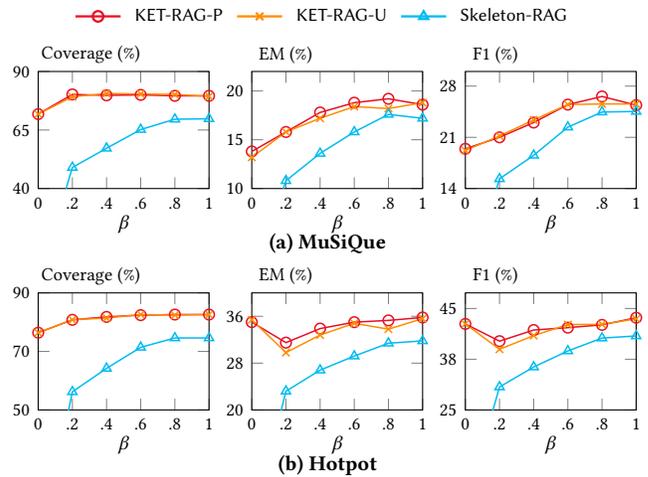
\begin{figure}[t]
\centering
\begin{tikzpicture}
    \begin{customlegend}[legend columns=4,
        legend entries={\sketragp, \sketragu, \skeletonrag\eat{, \skeletonragu}}
        ,
        legend style={at={(0.45,1.05)},anchor=north,draw=none,font=\footnotesize,column sep=0.1cm}]
    \addlegendimage{line width=0.3mm,mark size=2pt,mark=o,color=Red}
    \addlegendimage{line width=0.3mm,mark size=2pt,mark=x,color=Orange}
    \addlegendimage{line width=0.3mm,mark size=2pt,mark=triangle,color=LightBlue}
    \end{customlegend}
\end{tikzpicture}
\\
\subfloat[MuSiQue]{
\hspace{-1.2mm}\begin{tikzpicture}[scale=1]
    \begin{axis}[
        height=\columnwidth/2.7,
        width=\columnwidth/2.2,
        ylabel={Coverage (\%)},
        xlabel={$\beta$},
        xmin=0.0, xmax=1.0,
        ymin=0.4, ymax=0.9,
        xtick={0, 0.2, 0.4, 0.6, 0.8, 1.0},
        xticklabel style = {font=\footnotesize},
        xticklabels={0, .2, .4, .6, .8, 1},
        ytick={0.4, 0.65, 0.9},
        yticklabels={40, 65, 90},
        every axis y label/.style={at={(current axis.north west)},right=7mm,above=0mm},
        label style={font=\footnotesize},
        tick label style={font=\footnotesize},
        every axis x label/.style={at={(current axis.south)},right=0mm,above=-7mm},
        label style={font=\footnotesize},
        tick label style={font=\footnotesize},
    ]

    \addplot[line width=0.2mm,mark size=2pt,mark=o,color=Red]
        plot coordinates { 
(0.0, 0.718)
(0.2, 0.802)
(0.4, 0.798)
(0.6, 0.8)
(0.8, 0.796)
(1.0, 0.796)
    };

    \addplot[line width=0.2mm,mark size=2pt,mark=x,color=Orange]
        plot coordinates { 
(0.0, 0.718)
(0.2, 0.792)
(0.4, 0.806)
(0.6, 0.804)
(0.8, 0.802)
(1.0, 0.794)
    };

    \addplot[line width=0.2mm,mark size=2pt,mark=triangle,color=LightBlue]
        plot coordinates { 
(0.0, 0.0)
(0.2, 0.49)
(0.4, 0.572)
(0.6, 0.652)
(0.8, 0.696)
(1.0, 0.698)
    };


\end{axis}

\end{tikzpicture}
\hspace{-1.2mm}\begin{tikzpicture}[scale=1]
    \begin{axis}[
        height=\columnwidth/2.7,
        width=\columnwidth/2.2,
        ylabel={EM (\%)},
        xlabel={$\beta$},
        xmin=0.0, xmax=1.0,
        ymin=0.1, ymax=0.22,
        xtick={0, 0.2, 0.4, 0.6, 0.8, 1.0},
        xticklabel style = {font=\footnotesize},
        xticklabels={0, .2, .4, .6, .8, 1},
        ytick={0.1, 0.15, 0.2},
        yticklabels={10, 15, 20},
        every axis y label/.style={at={(current axis.north west)},right=4.5mm,above=0mm},
        label style={font=\footnotesize},
        tick label style={font=\footnotesize},
        every axis x label/.style={at={(current axis.south)},right=0mm,above=-7mm},
        label style={font=\footnotesize},
        tick label style={font=\footnotesize},
    ]

    \addplot[line width=0.2mm,mark size=2pt,mark=o,color=Red]
        plot coordinates { 
(0.0, 0.138)
(0.2, 0.158)
(0.4, 0.178)
(0.6, 0.188)
(0.8, 0.192)
(1.0, 0.186)
    };

    \addplot[line width=0.2mm,mark size=2pt,mark=x,color=Orange]
        plot coordinates { 
(0.0, 0.132)
(0.2, 0.158)
(0.4, 0.172)
(0.6, 0.184)
(0.8, 0.182)
(1.0, 0.188)
    };

    \addplot[line width=0.2mm,mark size=2pt,mark=triangle,color=LightBlue]
        plot coordinates { 
(0.0, 0.0)
(0.2, 0.108)
(0.4, 0.136)
(0.6, 0.158)
(0.8, 0.176)
(1.0, 0.172)
    };
    
    \end{axis}

\end{tikzpicture}
\hspace{-1.2mm}\begin{tikzpicture}[scale=1]
    \begin{axis}[
        height=\columnwidth/2.7,
        width=\columnwidth/2.2,
        ylabel={F1 (\%)},
        xlabel={$\beta$},
        xmin=0.0, xmax=1.0,
        ymin=0.14, ymax=0.3,
        xtick={0, 0.2, 0.4, 0.6, 0.8, 1.0},
        xticklabel style = {font=\footnotesize},
        xticklabels={0, .2, .4, .6, .8, 1},
        ytick={0.14, 0.21, 0.28},
        yticklabels={14, 21, 28},
        every axis y label/.style={at={(current axis.north west)},right=4mm,above=0mm},
        label style={font=\footnotesize},
        tick label style={font=\footnotesize},
        every axis x label/.style={at={(current axis.south)},right=0mm,above=-7mm},
        label style={font=\footnotesize},
        tick label style={font=\footnotesize},
    ]

    \addplot[line width=0.2mm,mark size=2pt,mark=o,color=Red]
        plot coordinates { 
(0.0, 0.194302287)
(0.2, 0.209992062)
(0.4, 0.23021343)
(0.6, 0.254704947)
(0.8, 0.265845586)
(1.0, 0.253952005)
    };

    \addplot[line width=0.2mm,mark size=2pt,mark=x,color=Orange]
        plot coordinates { 
(0.0, 0.192066627)
(0.2, 0.211880355)
(0.4, 0.233691958)
(0.6, 0.254856333)
(0.8, 0.255602241)
(1.0, 0.255616111)
    };

    \addplot[line width=0.2mm,mark size=2pt,mark=triangle,color=LightBlue]
        plot coordinates { 
(0.0, 0.0)
(0.2, 0.153266376)
(0.4, 0.185392483)
(0.6, 0.223946198)
(0.8, 0.244550107)
(1.0, 0.245370649)
    };

\end{axis}

\end{tikzpicture}
}%
\\
\subfloat[Hotpot]{
\hspace{-1.2mm}\begin{tikzpicture}[scale=1]
    \begin{axis}[
        height=\columnwidth/2.7,
        width=\columnwidth/2.2,
        ylabel={Coverage (\%)},
        xlabel={$\beta$},
        xmin=0.0, xmax=1.0,
        ymin=0.5, ymax=0.9,
        xtick={0, 0.2, 0.4, 0.6, 0.8, 1.0},
        xticklabel style = {font=\footnotesize},
        xticklabels={0, .2, .4, .6, .8, 1},
        ytick={0.5, 0.7, 0.9},
        yticklabels={50, 70, 90},
        every axis y label/.style={at={(current axis.north west)},right=7mm,above=0mm},
        label style={font=\footnotesize},
        tick label style={font=\footnotesize},
        every axis x label/.style={at={(current axis.south)},right=0mm,above=-7mm},
        label style={font=\footnotesize},
        tick label style={font=\footnotesize},
    ]

    \addplot[line width=0.2mm,mark size=2pt,mark=o,color=Red]
        plot coordinates { 
(0.0, 0.764)
(0.2, 0.808)
(0.4, 0.818)
(0.6, 0.824)
(0.8, 0.826)
(1.0, 0.826)
    };

    \addplot[line width=0.2mm,mark size=2pt,mark=x,color=Orange]
        plot coordinates { 
(0.0, 0.764)
(0.2, 0.808)
(0.4, 0.814)
(0.6, 0.826)
(0.8, 0.824)
(1.0, 0.826)
    };

    \addplot[line width=0.2mm,mark size=2pt,mark=triangle,color=LightBlue]
        plot coordinates { 
(0.0, 0.0)
(0.2, 0.562)
(0.4, 0.642)
(0.6, 0.714)
(0.8, 0.746)
(1.0, 0.746)
    };

    \end{axis}

\end{tikzpicture}
\hspace{-1.2mm}\begin{tikzpicture}[scale=1]
    \begin{axis}[
        height=\columnwidth/2.7,
        width=\columnwidth/2.2,
        ylabel={EM (\%)},
        xlabel={$\beta$},
        xmin=0.0, xmax=1.0,
        ymin=0.2, ymax=0.4,
        xtick={0, 0.2, 0.4, 0.6, 0.8, 1.0},
        xticklabel style = {font=\footnotesize},
        xticklabels={0, .2, .4, .6, .8, 1},
        ytick={0.2, 0.28, 0.36},
        yticklabels={20, 28, 36},
        every axis y label/.style={at={(current axis.north west)},right=4.5mm,above=0mm},
        label style={font=\footnotesize},
        tick label style={font=\footnotesize},
        every axis x label/.style={at={(current axis.south)},right=0mm,above=-7mm},
        label style={font=\footnotesize},
        tick label style={font=\footnotesize},
    ]

    \addplot[line width=0.2mm,mark size=2pt,mark=o,color=Red]
        plot coordinates { 
(0.0, 0.35)
(0.2, 0.315)
(0.4, 0.339)
(0.6, 0.35)
(0.8, 0.353)
(1.0, 0.358)
    };

    \addplot[line width=0.2mm,mark size=2pt,mark=x,color=Orange]
        plot coordinates { 
(0.0, 0.354)
(0.2, 0.298)
(0.4, 0.328)
(0.6, 0.348)
(0.8, 0.338)
(1.0, 0.356)
    };

    \addplot[line width=0.2mm,mark size=2pt,mark=triangle,color=LightBlue]
        plot coordinates { 
(0.0, 0.0)
(0.2, 0.232)
(0.4, 0.268)
(0.6, 0.292)
(0.8, 0.314)
(1.0, 0.318)
    };

    \end{axis}

\end{tikzpicture}
\hspace{-1.2mm}\begin{tikzpicture}[scale=1]
    \begin{axis}[
        height=\columnwidth/2.7,
        width=\columnwidth/2.2,
        ylabel={F1 (\%)},
        xlabel={$\beta$},
        xmin=0.0, xmax=1.0,
        ymin=0.25, ymax=0.55,
        xtick={0, 0.2, 0.4, 0.6, 0.8, 1.0},
        xticklabel style = {font=\footnotesize},
        xticklabels={0, .2, .4, .6, .8, 1},
        ytick={0.25,0.38, 0.51},
        yticklabels={25, 38, 45, 51},
        every axis y label/.style={at={(current axis.north west)},right=4mm,above=0mm},
        label style={font=\footnotesize},
        tick label style={font=\footnotesize},
        every axis x label/.style={at={(current axis.south)},right=0mm,above=-7mm},
        label style={font=\footnotesize},
        tick label style={font=\footnotesize},
    ]

    \addplot[line width=0.2mm,mark size=2pt,mark=o,color=Red]
        plot coordinates { 
(0.0, 0.470434704)
(0.2, 0.426276111)
(0.4, 0.454788870)
(0.6, 0.460886061)
(0.8, 0.468036694)
(1.0, 0.486306113)
    };

    \addplot[line width=0.2mm,mark size=2pt,mark=x,color=Orange]
        plot coordinates { 
(0.0, 0.470574266)
(0.2, 0.405430301)
(0.4, 0.44053814)
(0.6, 0.468784679)
(0.8, 0.469378725)
(1.0, 0.483362192)
    };

    \addplot[line width=0.2mm,mark size=2pt,mark=triangle,color=LightBlue]
        plot coordinates { 
(0.0, 0.0)
(0.2, 0.308885359)
(0.4, 0.360015033)
(0.6, 0.401194074)
(0.8, 0.434159393)
(1.0, 0.439605784)
    };

    \end{axis}

\end{tikzpicture}
}%
\vspace{-2mm}
\caption{Answer quality by varying $\beta$.} \label{fig:quality-beta}
\vspace{-4mm}
\end{figure}

%% file: figures/fig-theta.tex
\begin{figure}[t]
\centering
\begin{tikzpicture}
    \begin{customlegend}[legend columns=3,
        legend entries={\sketragp, \sketragu, \hybridrag}
        ,
        legend style={at={(0.45,1.05)},anchor=north,draw=none,font=\footnotesize,column sep=0.1cm}]
    \addlegendimage{line width=0.3mm,mark size=2pt,mark=o,color=Red}
    \addlegendimage{line width=0.3mm,mark size=2pt,mark=x,color=Orange}
    \addlegendimage{line width=0.3mm,mark size=2pt,mark=pentagon,color=Green}
    \end{customlegend}
\end{tikzpicture}
\\
\subfloat[MuSiQue]{
\hspace{-1.2mm}\begin{tikzpicture}[scale=1]
    \begin{axis}[
        height=\columnwidth/2.7,
        width=\columnwidth/2.2,
        ylabel={Coverage (\%)},
        xlabel={$\theta$},
        xmin=0.0, xmax=1.0,
        ymin=0.65, ymax=0.85,
        xtick={0, 0.2, 0.4, 0.6, 0.8, 1.0},
        xticklabel style = {font=\footnotesize},
        xticklabels={0, .2, .4, .6, .8, 1},
        ytick={0.65,0.75,0.85},
        yticklabels={65,75,85},
        every axis y label/.style={at={(current axis.north west)},right=7mm,above=0mm},
        label style={font=\footnotesize},
        tick label style={font=\footnotesize},
        every axis x label/.style={at={(current axis.south)},right=0mm,above=-7mm},
        label style={font=\footnotesize},
        tick label style={font=\footnotesize},
    ]

    \addplot[line width=0.2mm,mark size=2pt,mark=o,color=Red]
        plot coordinates { 
(0.0, 0.782)
(0.2, 0.808)
(0.4, 0.796)
(0.6, 0.788)
(0.8, 0.752)
(1.0, 0.696)
    };

    \addplot[line width=0.2mm,mark size=2pt,mark=x,color=Orange]
        plot coordinates { 
(0.0, 0.782)
(0.2, 0.812)
(0.4, 0.802)
(0.6, 0.794)
(0.8, 0.758)
(1.0, 0.71)
    };

    \addplot[line width=0.2mm,mark size=2pt,mark=pentagon,color=Green]
        plot coordinates { 
(0.0, 0.768)
(0.2, 0.816)
(0.4, 0.804)
(0.6, 0.786)
(0.8, 0.76)
(1.0, 0.696)
    };

\end{axis}

\end{tikzpicture}
\hspace{-1.2mm}\begin{tikzpicture}[scale=1]
    \begin{axis}[
        height=\columnwidth/2.7,
        width=\columnwidth/2.2,
        ylabel={EM (\%)},
        xlabel={$\theta$},
        xmin=0.0, xmax=1.0,
        ymin=0.1, ymax=0.22,
        xtick={0, 0.2, 0.4, 0.6, 0.8, 1.0},
        xticklabel style = {font=\footnotesize},
        xticklabels={0, .2, .4, .6, .8, 1},
        ytick={0.10, 0.15,0.2},
        yticklabels={10, 15, 20},
        every axis y label/.style={at={(current axis.north west)},right=4.5mm,above=0mm},
        label style={font=\footnotesize},
        tick label style={font=\footnotesize},
        every axis x label/.style={at={(current axis.south)},right=0mm,above=-7mm},
        label style={font=\footnotesize},
        tick label style={font=\footnotesize},
    ]

    \addplot[line width=0.2mm,mark size=2pt,mark=o,color=Red]
        plot coordinates { 
(0, 0.146)
(0.2, 0.194)
(0.4, 0.186)
(0.6, 0.188)
(0.8, 0.184)
(1.0, 0.174)
    };

    \addplot[line width=0.2mm,mark size=2pt,mark=x,color=Orange]
        plot coordinates { 
(0.0, 0.142)
(0.2, 0.182)
(0.4, 0.182)
(0.6, 0.18)
(0.8, 0.182)
(1.0, 0.17)
    };

    \addplot[line width=0.2mm,mark size=2pt,mark=pentagon,color=Green]
        plot coordinates { 
(0.0, 0.12)
(0.2, 0.18)
(0.4, 0.178)
(0.6, 0.19)
(0.8, 0.18)
(1.0, 0.17)
    };
    
    \end{axis}

\end{tikzpicture}
\hspace{-1.2mm}\begin{tikzpicture}[scale=1]
    \begin{axis}[
        height=\columnwidth/2.7,
        width=\columnwidth/2.2,
        ylabel={F1 (\%)},
        xlabel={$\theta$},
        xmin=0.0, xmax=1.0,
        ymin=0.18, ymax=0.28,
        xtick={0, 0.2, 0.4, 0.6, 0.8, 1.0},
        xticklabel style = {font=\footnotesize},
        xticklabels={0, .2, .4, .6, .8, 1},
        ytick={0.18,0.23,0.28},
        yticklabels={18,23,28},
        every axis y label/.style={at={(current axis.north west)},right=4mm,above=0mm},
        label style={font=\footnotesize},
        tick label style={font=\footnotesize},
        every axis x label/.style={at={(current axis.south)},right=0mm,above=-7mm},
        label style={font=\footnotesize},
        tick label style={font=\footnotesize},
    ]

    \addplot[line width=0.2mm,mark size=2pt,mark=o,color=Red]
        plot coordinates { 
(0, 0.205702525)
(0.2, 0.261869012)
(0.4, 0.253952005)
(0.6, 0.262738242)
(0.8, 0.251539304)
(1.0, 0.25075364)
    };

    \addplot[line width=0.2mm,mark size=2pt,mark=x,color=Orange]
        plot coordinates { 
(0.0, 0.204517181)
(0.2, 0.251324036)
(0.4, 0.255602241)
(0.6, 0.254754718)
(0.8, 0.254316549)
(1.0, 0.238788916)
    };

    \addplot[line width=0.2mm,mark size=2pt,mark=pentagon,color=Green]
        plot coordinates { 
(0.0, 0.186808122)
(0.2, 0.243938036)
(0.4, 0.247127927)
(0.6, 0.256951852)
(0.8, 0.255521173)
(1.0, 0.243875902)
    };

\end{axis}

\end{tikzpicture}
}%
\\
\subfloat[Hotpot]{
\hspace{-1.2mm}\begin{tikzpicture}[scale=1]
    \begin{axis}[
        height=\columnwidth/2.7,
        width=\columnwidth/2.2,
        ylabel={Coverage (\%)},
        xlabel={$\theta$},
        xmin=0.0, xmax=1.0,
        ymin=0.7, ymax=0.9,
        xtick={0, 0.2, 0.4, 0.6, 0.8, 1.0},
        xticklabel style = {font=\footnotesize},
        xticklabels={0, .2, .4, .6, .8, 1},
        ytick={0.7,0.8,0.9},
        yticklabels={70,80,90},
        every axis y label/.style={at={(current axis.north west)},right=7mm,above=0mm},
        label style={font=\footnotesize},
        tick label style={font=\footnotesize},
        every axis x label/.style={at={(current axis.south)},right=0mm,above=-7mm},
        label style={font=\footnotesize},
        tick label style={font=\footnotesize},
    ]

    \addplot[line width=0.2mm,mark size=2pt,mark=pentagon,color=Green]
        plot coordinates { 
(0.0, 0.74)
(0.2, 0.818)
(0.4, 0.81)
(0.6, 0.792)
(0.8, 0.774)
(1.0, 0.746)
    };

    \addplot[line width=0.2mm,mark size=2pt,mark=x,color=DeepBlue]
        plot coordinates { 

    };

    \addplot[line width=0.2mm,mark size=2pt,mark=x,color=Orange]
        plot coordinates { 
(0.0, 0.824)
(0.2, 0.852)
(0.4, 0.824)
(0.6, 0.802)
(0.8, 0.78)
(1.0, 0.736)
    };

    \addplot[line width=0.2mm,mark size=2pt,mark=o,color=Red]
        plot coordinates { 
(0.0, 0.822)
(0.2, 0.846)
(0.4, 0.826)
(0.6, 0.804)
(0.8, 0.78)
(1.0, 0.746)
    };

    \end{axis}

\end{tikzpicture}
\hspace{-1.2mm}\begin{tikzpicture}[scale=1]
    \begin{axis}[
        height=\columnwidth/2.7,
        width=\columnwidth/2.2,
        ylabel={EM (\%)},
        xlabel={$\theta$},
        xmin=0.0, xmax=1.0,
        ymin=0.28, ymax=0.38,
        xtick={0, 0.2, 0.4, 0.6, 0.8, 1.0},
        xticklabel style = {font=\footnotesize},
        xticklabels={0, .2, .4, .6, .8, 1},
        ytick={0.30, 0.33, 0.36},
        yticklabels={30, 33, 36},
        every axis y label/.style={at={(current axis.north west)},right=4.5mm,above=0mm},
        label style={font=\footnotesize},
        tick label style={font=\footnotesize},
        every axis x label/.style={at={(current axis.south)},right=0mm,above=-7mm},
        label style={font=\footnotesize},
        tick label style={font=\footnotesize},
    ]

    \addplot[line width=0.2mm,mark size=2pt,mark=pentagon,color=Green]
        plot coordinates { 
(0.0, 0.332)
(0.2, 0.34)
(0.4, 0.34)
(0.6, 0.33)
(0.8, 0.334)
(1.0, 0.312)
    };

    \addplot[line width=0.2mm,mark size=2pt,mark=o,color=Red]
        plot coordinates { 
(0, 0.35)
(0.2, 0.356)
(0.4, 0.358)
(0.6, 0.34)
(0.8, 0.326)
(1.0, 0.31)
    };

    \addplot[line width=0.2mm,mark size=2pt,mark=x,color=Orange]
        plot coordinates { 
(0.0, 0.348)
(0.2, 0.348)
(0.4, 0.348)
(0.6, 0.352)
(0.8, 0.350)
(1.0, 0.316)
    };
    
    \end{axis}

\end{tikzpicture}
\hspace{-1.2mm}\begin{tikzpicture}[scale=1]
    \begin{axis}[
        height=\columnwidth/2.7,
        width=\columnwidth/2.2,
        ylabel={F1 (\%)},
        xlabel={$\theta$},
        xmin=0.0, xmax=1.0,
        ymin=0.4, ymax=0.5,
        xtick={0, 0.2, 0.4, 0.6, 0.8, 1.0},
        xticklabel style = {font=\footnotesize},
        xticklabels={0, .2, .4, .6, .8, 1},
        ytick={0.4, 0.45, 0.5},
        yticklabels={40, 45, 50},
        every axis y label/.style={at={(current axis.north west)},right=4mm,above=0mm},
        label style={font=\footnotesize},
        tick label style={font=\footnotesize},
        every axis x label/.style={at={(current axis.south)},right=0mm,above=-7mm},
        label style={font=\footnotesize},
        tick label style={font=\footnotesize},
    ]

    \addplot[line width=0.2mm,mark size=2pt,mark=pentagon,color=Green]
        plot coordinates { 
(0.0, 0.437930177)
(0.2, 0.462061217)
(0.4, 0.458510645)
(0.6, 0.453968044)
(0.8, 0.453729208)
(1.0, 0.431479231)
    };

    \addplot[line width=0.2mm,mark size=2pt,mark=o,color=Red]
        plot coordinates { 
(0, 0.470963143)
(0.2, 0.484303943)
(0.4, 0.486306113)
(0.6, 0.472271351)
(0.8, 0.450982308)
(1.0, 0.429801034)
    };

    \addplot[line width=0.2mm,mark size=2pt,mark=x,color=Orange]
        plot coordinates { 
(0.0, 0.469989276)
(0.2, 0.475449933)
(0.4, 0.4751002065)
(0.6, 0.4719215465)
(0.8, 0.4705509015)
(1.0, 0.426568579)
    };

    \end{axis}

\end{tikzpicture}
}%
\vspace{-2mm}
\caption{Answer quality by varying $\theta$.} \label{fig:quality-theta}
\vspace{-2mm}
\end{figure}
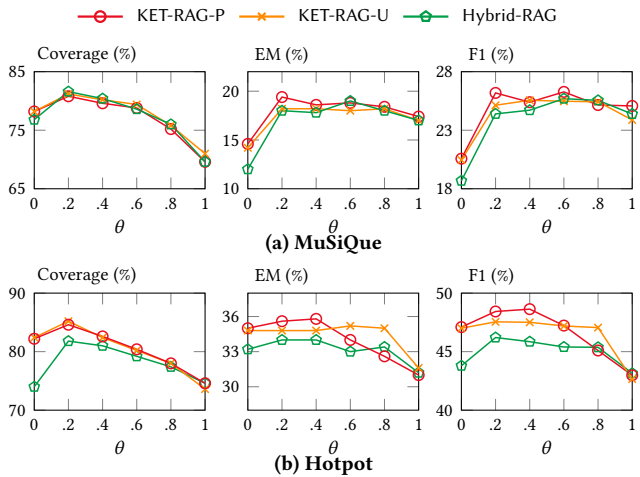

%% file: sections/conclusion.tex
\section{Conclusions}
In this work, we propose \sketrag, a cost-efficient multi-granular indexing framework for Graph-RAG systems. By integrating a knowledge graph skeleton with a text-keyword bipartite graph, \sketrag improves retrieval and generation quality while significantly reducing indexing costs. Our approach matches or surpasses Microsoft's \graphrag in retrieval quality while reducing indexing costs by over an order of magnitude, and improves generation quality by up to 32.4\% with 20\% lower indexing costs.
For future work, we plan to extend \sketrag to global search scenarios and explore adaptive paradigms for real-world scalable deployment.

%% file: main-tr.bbl

\begin{thebibliography}{39}


\ifx \showCODEN    \undefined \def \showCODEN     #1{\unskip}     \fi
\ifx \showISBNx    \undefined \def \showISBNx     #1{\unskip}     \fi
\ifx \showISBNxiii \undefined \def \showISBNxiii  #1{\unskip}     \fi
\ifx \showISSN     \undefined \def \showISSN      #1{\unskip}     \fi
\ifx \showLCCN     \undefined \def \showLCCN      #1{\unskip}     \fi
\ifx \shownote     \undefined \def \shownote      #1{#1}          \fi
\ifx \showarticletitle \undefined \def \showarticletitle #1{#1}   \fi
\ifx \showURL      \undefined \def \showURL       {\relax}        \fi
\providecommand\bibfield[2]{#2}
\providecommand\bibinfo[2]{#2}
\providecommand\natexlab[1]{#1}
\providecommand\showeprint[2][]{arXiv:#2}

\bibitem[Alhanahnah et~al\mbox{.}(2024)]%
        {alhanahnah2024depesrag}
\bibfield{author}{\bibinfo{person}{Mohannad Alhanahnah}, \bibinfo{person}{Yazan Boshmaf}, {and} \bibinfo{person}{Benoit Baudry}.} \bibinfo{year}{2024}\natexlab{}.
\newblock \showarticletitle{DepesRAG: Towards Managing Software Dependencies using Large Language Models}.
\newblock \bibinfo{journal}{\emph{arXiv preprint arXiv:2405.20455}} (\bibinfo{year}{2024}).
\newblock


\bibitem[Arnold and Romero(2022)]%
        {arnold2022ediscovery}
\bibfield{author}{\bibinfo{person}{Shawn Arnold} {and} \bibinfo{person}{Clayton Romero}.} \bibinfo{year}{2022}\natexlab{}.
\newblock \bibinfo{title}{The Vital Role of Managing e-Discovery}.
\newblock
\urldef\tempurl%
\url{https://legal-tech.blog/the-vital-role-of-managing-e-discovery}
\showURL{%
\tempurl}


\bibitem[Cao et~al\mbox{.}(2024)]%
        {cao24companykg}
\bibfield{author}{\bibinfo{person}{Lele Cao}, \bibinfo{person}{Vilhelm von Ehrenheim}, \bibinfo{person}{Mark Granroth-Wilding}, \bibinfo{person}{Richard Anselmo~Stahl}, \bibinfo{person}{Andrew McCornack}, \bibinfo{person}{Armin Catovic}, {and} \bibinfo{person}{Dhiana~Deva Cavalcanti~Rocha}.} \bibinfo{year}{2024}\natexlab{}.
\newblock \showarticletitle{CompanyKG: A Large-Scale Heterogeneous Graph for Company Similarity Quantification}. In \bibinfo{booktitle}{\emph{Proceedings of the 30th ACM SIGKDD Conference on Knowledge Discovery and Data Mining}}. \bibinfo{pages}{4816–4827}.
\newblock


\bibitem[Chen et~al\mbox{.}(2024)]%
        {chen24rarebench}
\bibfield{author}{\bibinfo{person}{Xuanzhong Chen}, \bibinfo{person}{Xiaohao Mao}, \bibinfo{person}{Qihan Guo}, \bibinfo{person}{Lun Wang}, \bibinfo{person}{Shuyang Zhang}, {and} \bibinfo{person}{Ting Chen}.} \bibinfo{year}{2024}\natexlab{}.
\newblock \showarticletitle{RareBench: Can LLMs Serve as Rare Diseases Specialists?}. In \bibinfo{booktitle}{\emph{Proceedings of the 30th ACM SIGKDD Conference on Knowledge Discovery and Data Mining}}. \bibinfo{pages}{4850–4861}.
\newblock


\bibitem[Colombo(2024)]%
        {colombo2024leveraging}
\bibfield{author}{\bibinfo{person}{Andrea Colombo}.} \bibinfo{year}{2024}\natexlab{}.
\newblock \showarticletitle{Leveraging Knowledge Graphs and LLMs to Support and Monitor Legislative Systems}. In \bibinfo{booktitle}{\emph{Proceedings of the 33rd ACM International Conference on Information and Knowledge Management}}. \bibinfo{pages}{5443--5446}.
\newblock


\bibitem[Dehghan et~al\mbox{.}(2024)]%
        {dehghan2024ewek}
\bibfield{author}{\bibinfo{person}{Mohammad Dehghan}, \bibinfo{person}{Mohammad Alomrani}, \bibinfo{person}{Sunyam Bagga}, \bibinfo{person}{David Alfonso-Hermelo}, \bibinfo{person}{Khalil Bibi}, \bibinfo{person}{Abbas Ghaddar}, \bibinfo{person}{Yingxue Zhang}, \bibinfo{person}{Xiaoguang Li}, \bibinfo{person}{Jianye Hao}, \bibinfo{person}{Qun Liu}, \bibinfo{person}{Jimmy Lin}, \bibinfo{person}{Boxing Chen}, \bibinfo{person}{Prasanna Parthasarathi}, \bibinfo{person}{Mahdi Biparva}, {and} \bibinfo{person}{Mehdi Rezagholizadeh}.} \bibinfo{year}{2024}\natexlab{}.
\newblock \showarticletitle{{EWEK}-{QA} : Enhanced Web and Efficient Knowledge Graph Retrieval for Citation-based Question Answering Systems}. In \bibinfo{booktitle}{\emph{Proceedings of the 62nd Annual Meeting of the Association for Computational Linguistics (Volume 1: Long Papers)}}.
\newblock


\bibitem[Delile et~al\mbox{.}(2024)]%
        {delile2024graph}
\bibfield{author}{\bibinfo{person}{Julien Delile}, \bibinfo{person}{Srayanta Mukherjee}, \bibinfo{person}{Anton Van~Pamel}, {and} \bibinfo{person}{Leonid Zhukov}.} \bibinfo{year}{2024}\natexlab{}.
\newblock \showarticletitle{Graph-Based Retriever Captures the Long Tail of Biomedical Knowledge}.
\newblock \bibinfo{journal}{\emph{arXiv preprint arXiv:2402.12352}} (\bibinfo{year}{2024}).
\newblock


\bibitem[Edge et~al\mbox{.}(2024)]%
        {edge2024local}
\bibfield{author}{\bibinfo{person}{Darren Edge}, \bibinfo{person}{Ha Trinh}, \bibinfo{person}{Newman Cheng}, \bibinfo{person}{Joshua Bradley}, \bibinfo{person}{Alex Chao}, \bibinfo{person}{Apurva Mody}, \bibinfo{person}{Steven Truitt}, {and} \bibinfo{person}{Jonathan Larson}.} \bibinfo{year}{2024}\natexlab{}.
\newblock \showarticletitle{From local to global: A graph rag approach to query-focused summarization}.
\newblock \bibinfo{journal}{\emph{arXiv preprint arXiv:2404.16130}} (\bibinfo{year}{2024}).
\newblock


\bibitem[Fan et~al\mbox{.}(2024)]%
        {fan24survey}
\bibfield{author}{\bibinfo{person}{Wenqi Fan}, \bibinfo{person}{Yujuan Ding}, \bibinfo{person}{Liangbo Ning}, \bibinfo{person}{Shijie Wang}, \bibinfo{person}{Hengyun Li}, \bibinfo{person}{Dawei Yin}, \bibinfo{person}{Tat-Seng Chua}, {and} \bibinfo{person}{Qing Li}.} \bibinfo{year}{2024}\natexlab{}.
\newblock \showarticletitle{A Survey on RAG Meeting LLMs: Towards Retrieval-Augmented Large Language Models}. In \bibinfo{booktitle}{\emph{Proceedings of the 30th ACM SIGKDD Conference on Knowledge Discovery and Data Mining}}. \bibinfo{pages}{6491–6501}.
\newblock


\bibitem[Gao et~al\mbox{.}(2023)]%
        {gao2023precise}
\bibfield{author}{\bibinfo{person}{Luyu Gao}, \bibinfo{person}{Xueguang Ma}, \bibinfo{person}{Jimmy Lin}, {and} \bibinfo{person}{Jamie Callan}.} \bibinfo{year}{2023}\natexlab{}.
\newblock \showarticletitle{Precise zero-shot dense retrieval without relevance labels}. In \bibinfo{booktitle}{\emph{Proceedings of the 61st Annual Meeting of the Association for Computational Linguistics (Volume 1: Long Papers)}}. \bibinfo{pages}{1762--1777}.
\newblock


\bibitem[Group and OpenKG(2023)]%
        {antgrouprag}
\bibfield{author}{\bibinfo{person}{Ant Group} {and} \bibinfo{person}{OpenKG}.} \bibinfo{year}{2023}\natexlab{}.
\newblock \bibinfo{title}{Semantic-enhanced Programmable Knowledge Graph (SPG) White paper (v1.0)}.
\newblock \bibinfo{howpublished}{\url{https://spg.openkg.cn/en-US}}.
\newblock


\bibitem[Gu et~al\mbox{.}(2021)]%
        {gu2021domain}
\bibfield{author}{\bibinfo{person}{Yu Gu}, \bibinfo{person}{Robert Tinn}, \bibinfo{person}{Hao Cheng}, \bibinfo{person}{Michael Lucas}, \bibinfo{person}{Naoto Usuyama}, \bibinfo{person}{Xiaodong Liu}, \bibinfo{person}{Tristan Naumann}, \bibinfo{person}{Jianfeng Gao}, {and} \bibinfo{person}{Hoifung Poon}.} \bibinfo{year}{2021}\natexlab{}.
\newblock \showarticletitle{Domain-specific language model pretraining for biomedical natural language processing}.
\newblock \bibinfo{journal}{\emph{ACM Transactions on Computing for Healthcare (HEALTH)}} \bibinfo{volume}{3}, \bibinfo{number}{1} (\bibinfo{year}{2021}), \bibinfo{pages}{1--23}.
\newblock


\bibitem[Guo et~al\mbox{.}(2025)]%
        {guo2025lightrag}
\bibfield{author}{\bibinfo{person}{Zirui Guo}, \bibinfo{person}{Lianghao Xia}, \bibinfo{person}{Yanhua Yu}, \bibinfo{person}{Tu Ao}, {and} \bibinfo{person}{Chao Huang}.} \bibinfo{year}{2025}\natexlab{}.
\newblock \bibinfo{title}{LightRAG: Simple and Fast Retrieval-Augmented Generation}.
\newblock
\showeprint[arxiv]{2410.05779}~[cs.IR]
\urldef\tempurl%
\url{https://arxiv.org/abs/2410.05779}
\showURL{%
\tempurl}


\bibitem[Gutierrez et~al\mbox{.}(2024)]%
        {gutierrez2024hipporag}
\bibfield{author}{\bibinfo{person}{Bernal~Jimenez Gutierrez}, \bibinfo{person}{Yiheng Shu}, \bibinfo{person}{Yu Gu}, \bibinfo{person}{Michihiro Yasunaga}, {and} \bibinfo{person}{Yu Su}.} \bibinfo{year}{2024}\natexlab{}.
\newblock \showarticletitle{Hippo{RAG}: Neurobiologically Inspired Long-Term Memory for Large Language Models}. In \bibinfo{booktitle}{\emph{The Thirty-eighth Annual Conference on Neural Information Processing Systems}}.
\newblock
\urldef\tempurl%
\url{https://openreview.net/forum?id=hkujvAPVsg}
\showURL{%
\tempurl}


\bibitem[Han et~al\mbox{.}(2024)]%
        {han-etal-2024-rag}
\bibfield{author}{\bibinfo{person}{Rujun Han}, \bibinfo{person}{Yuhao Zhang}, \bibinfo{person}{Peng Qi}, \bibinfo{person}{Yumo Xu}, \bibinfo{person}{Jenyuan Wang}, \bibinfo{person}{Lan Liu}, \bibinfo{person}{William~Yang Wang}, \bibinfo{person}{Bonan Min}, {and} \bibinfo{person}{Vittorio Castelli}.} \bibinfo{year}{2024}\natexlab{}.
\newblock \showarticletitle{{RAG}-{QA} Arena: Evaluating Domain Robustness for Long-form Retrieval Augmented Question Answering}. In \bibinfo{booktitle}{\emph{Proceedings of the 2024 Conference on Empirical Methods in Natural Language Processing}}, \bibfield{editor}{\bibinfo{person}{Yaser Al-Onaizan}, \bibinfo{person}{Mohit Bansal}, {and} \bibinfo{person}{Yun-Nung Chen}} (Eds.). \bibinfo{publisher}{Association for Computational Linguistics}, \bibinfo{address}{Miami, Florida, USA}, \bibinfo{pages}{4354--4374}.
\newblock
\href{https://doi.org/10.18653/v1/2024.emnlp-main.249}{doi:\nolinkurl{10.18653/v1/2024.emnlp-main.249}}


\bibitem[He et~al\mbox{.}(2024)]%
        {he2024g}
\bibfield{author}{\bibinfo{person}{Xiaoxin He}, \bibinfo{person}{Yijun Tian}, \bibinfo{person}{Yifei Sun}, \bibinfo{person}{Nitesh~V Chawla}, \bibinfo{person}{Thomas Laurent}, \bibinfo{person}{Yann LeCun}, \bibinfo{person}{Xavier Bresson}, {and} \bibinfo{person}{Bryan Hooi}.} \bibinfo{year}{2024}\natexlab{}.
\newblock \showarticletitle{G-retriever: Retrieval-augmented generation for textual graph understanding and question answering}. In \bibinfo{booktitle}{\emph{The Thirty-eighth Annual Conference on Neural Information Processing Systems}}.
\newblock


\bibitem[Jin et~al\mbox{.}(2024)]%
        {jin2024graph}
\bibfield{author}{\bibinfo{person}{Bowen Jin}, \bibinfo{person}{Chulin Xie}, \bibinfo{person}{Jiawei Zhang}, \bibinfo{person}{Kashob~Kumar Roy}, \bibinfo{person}{Yu Zhang}, \bibinfo{person}{Zheng Li}, \bibinfo{person}{Ruirui Li}, \bibinfo{person}{Xianfeng Tang}, \bibinfo{person}{Suhang Wang}, \bibinfo{person}{Yu Meng}, {et~al\mbox{.}}} \bibinfo{year}{2024}\natexlab{}.
\newblock \showarticletitle{Graph chain-of-thought: Augmenting large language models by reasoning on graphs}.
\newblock \bibinfo{journal}{\emph{arXiv preprint arXiv:2404.07103}} (\bibinfo{year}{2024}).
\newblock


\bibitem[Kalra et~al\mbox{.}(2024)]%
        {kalra2024hypa}
\bibfield{author}{\bibinfo{person}{Rishi Kalra}, \bibinfo{person}{Zekun Wu}, \bibinfo{person}{Ayesha Gulley}, \bibinfo{person}{Airlie Hilliard}, \bibinfo{person}{Xin Guan}, \bibinfo{person}{Adriano Koshiyama}, {and} \bibinfo{person}{Philip Treleaven}.} \bibinfo{year}{2024}\natexlab{}.
\newblock \showarticletitle{HyPA-RAG: A Hybrid Parameter Adaptive Retrieval-Augmented Generation System for AI Legal and Policy Applications}. In \bibinfo{booktitle}{\emph{Proceedings of the 1st Workshop on Customizable NLP: Progress and Challenges in Customizing NLP for a Domain, Application, Group, or Individual (CustomNLP4U)}}.
\newblock


\bibitem[Lewis et~al\mbox{.}(2020)]%
        {lewis2020retrieval}
\bibfield{author}{\bibinfo{person}{Patrick Lewis}, \bibinfo{person}{Ethan Perez}, \bibinfo{person}{Aleksandra Piktus}, \bibinfo{person}{Fabio Petroni}, \bibinfo{person}{Vladimir Karpukhin}, \bibinfo{person}{Naman Goyal}, \bibinfo{person}{Heinrich K{\"u}ttler}, \bibinfo{person}{Mike Lewis}, \bibinfo{person}{Wen-tau Yih}, \bibinfo{person}{Tim Rockt{\"a}schel}, {et~al\mbox{.}}} \bibinfo{year}{2020}\natexlab{}.
\newblock \showarticletitle{Retrieval-augmented generation for knowledge-intensive nlp tasks}.
\newblock \bibinfo{journal}{\emph{Advances in Neural Information Processing Systems}}  \bibinfo{volume}{33} (\bibinfo{year}{2020}), \bibinfo{pages}{9459--9474}.
\newblock


\bibitem[Li et~al\mbox{.}(2024b)]%
        {li2024dalk}
\bibfield{author}{\bibinfo{person}{Dawei Li}, \bibinfo{person}{Shu Yang}, \bibinfo{person}{Zhen Tan}, \bibinfo{person}{Jae~Young Baik}, \bibinfo{person}{Sukwon Yun}, \bibinfo{person}{Joseph Lee}, \bibinfo{person}{Aaron Chacko}, \bibinfo{person}{Bojian Hou}, \bibinfo{person}{Duy Duong-Tran}, \bibinfo{person}{Ying Ding}, \bibinfo{person}{Huan Liu}, \bibinfo{person}{Li Shen}, {and} \bibinfo{person}{Tianlong Chen}.} \bibinfo{year}{2024}\natexlab{b}.
\newblock \showarticletitle{{DALK}: Dynamic Co-Augmentation of {LLM}s and {KG} to answer {A}lzheimer`s Disease Questions with Scientific Literature}. In \bibinfo{booktitle}{\emph{Findings of the Association for Computational Linguistics: EMNLP 2024}}. \bibinfo{pages}{2187--2205}.
\newblock


\bibitem[Li et~al\mbox{.}(2024a)]%
        {li2024graph}
\bibfield{author}{\bibinfo{person}{Zijian Li}, \bibinfo{person}{Qingyan Guo}, \bibinfo{person}{Jiawei Shao}, \bibinfo{person}{Lei Song}, \bibinfo{person}{Jiang Bian}, \bibinfo{person}{Jun Zhang}, {and} \bibinfo{person}{Rui Wang}.} \bibinfo{year}{2024}\natexlab{a}.
\newblock \showarticletitle{Graph Neural Network Enhanced Retrieval for Question Answering of LLMs}.
\newblock \bibinfo{journal}{\emph{arXiv preprint arXiv:2406.06572}} (\bibinfo{year}{2024}).
\newblock


\bibitem[Mavromatis and Karypis(2024)]%
        {mavromatis2024gnn}
\bibfield{author}{\bibinfo{person}{Costas Mavromatis} {and} \bibinfo{person}{George Karypis}.} \bibinfo{year}{2024}\natexlab{}.
\newblock \showarticletitle{GNN-RAG: Graph Neural Retrieval for Large Language Model Reasoning}.
\newblock \bibinfo{journal}{\emph{arXiv preprint arXiv:2405.20139}} (\bibinfo{year}{2024}).
\newblock


\bibitem[Min et~al\mbox{.}(2019)]%
        {min2019knowledge}
\bibfield{author}{\bibinfo{person}{Sewon Min}, \bibinfo{person}{Danqi Chen}, \bibinfo{person}{Luke Zettlemoyer}, {and} \bibinfo{person}{Hannaneh Hajishirzi}.} \bibinfo{year}{2019}\natexlab{}.
\newblock \showarticletitle{Knowledge guided text retrieval and reading for open domain question answering}.
\newblock \bibinfo{journal}{\emph{arXiv preprint arXiv:1911.03868}} (\bibinfo{year}{2019}).
\newblock


\bibitem[Mou et~al\mbox{.}(2024)]%
        {mou2024unifying}
\bibfield{author}{\bibinfo{person}{Xinyi Mou}, \bibinfo{person}{Zejun Li}, \bibinfo{person}{Hanjia Lyu}, \bibinfo{person}{Jiebo Luo}, {and} \bibinfo{person}{Zhongyu Wei}.} \bibinfo{year}{2024}\natexlab{}.
\newblock \showarticletitle{Unifying Local and Global Knowledge: Empowering Large Language Models as Political Experts with Knowledge Graphs}. In \bibinfo{booktitle}{\emph{Proceedings of the ACM on Web Conference 2024}}. \bibinfo{pages}{2603--2614}.
\newblock


\bibitem[Munikoti et~al\mbox{.}(2024)]%
        {munikoti2023atlantic}
\bibfield{author}{\bibinfo{person}{Sai Munikoti}, \bibinfo{person}{Anurag Acharya}, \bibinfo{person}{Sridevi Wagle}, {and} \bibinfo{person}{Sameera Horawalavithana}.} \bibinfo{year}{2024}\natexlab{}.
\newblock \showarticletitle{ATLANTIC: Structure-Aware Retrieval-Augmented Language Model for Interdisciplinary Science}.
\newblock \bibinfo{journal}{\emph{Proceedings of the Workshop on AI to Accelerate Science and Engineering (AI2ASE). Held in conjunction with the 38th AAAI Conference on Artificial Intelligence.}} (\bibinfo{year}{2024}).
\newblock


\bibitem[NebulaGraph(2023)]%
        {nebularag}
\bibfield{author}{\bibinfo{person}{NebulaGraph}.} \bibinfo{year}{2023}\natexlab{}.
\newblock \bibinfo{title}{NebulaGraph Launches Industry-First Graph RAG: Retrieval-Augmented Generation with LLM Based on Knowledge Graphs}.
\newblock \bibinfo{howpublished}{\url{https://www.nebula-graph.io/posts/graph-RAG}}.
\newblock


\bibitem[Neo4j(2023)]%
        {neo4jrag}
\bibfield{author}{\bibinfo{person}{Neo4j}.} \bibinfo{year}{2023}\natexlab{}.
\newblock \bibinfo{title}{NaLLM}.
\newblock \bibinfo{howpublished}{\url{https://github.com/neo4j/NaLLM}}.
\newblock


\bibitem[Page et~al\mbox{.}(1999)]%
        {page1999pagerank}
\bibfield{author}{\bibinfo{person}{Lawrence Page}, \bibinfo{person}{Sergey Brin}, \bibinfo{person}{Rajeev Motwani}, {and} \bibinfo{person}{Terry Winograd}.} \bibinfo{year}{1999}\natexlab{}.
\newblock \bibinfo{booktitle}{\emph{The PageRank citation ranking: Bringing order to the web.}}
\newblock \bibinfo{type}{{T}echnical {R}eport}. \bibinfo{institution}{Stanford InfoLab}.
\newblock


\bibitem[Peng et~al\mbox{.}(2024)]%
        {peng2024graph}
\bibfield{author}{\bibinfo{person}{Boci Peng}, \bibinfo{person}{Yun Zhu}, \bibinfo{person}{Yongchao Liu}, \bibinfo{person}{Xiaohe Bo}, \bibinfo{person}{Haizhou Shi}, \bibinfo{person}{Chuntao Hong}, \bibinfo{person}{Yan Zhang}, {and} \bibinfo{person}{Siliang Tang}.} \bibinfo{year}{2024}\natexlab{}.
\newblock \showarticletitle{Graph Retrieval-Augmented Generation: A Survey}.
\newblock \bibinfo{journal}{\emph{arXiv preprint arXiv:2408.08921}} (\bibinfo{year}{2024}).
\newblock


\bibitem[Peng and Yang(2024)]%
        {peng2024connecting}
\bibfield{author}{\bibinfo{person}{Zhuoyi Peng} {and} \bibinfo{person}{Yi Yang}.} \bibinfo{year}{2024}\natexlab{}.
\newblock \showarticletitle{Connecting the Dots: Inferring Patent Phrase Similarity with Retrieved Phrase Graphs}. In \bibinfo{booktitle}{\emph{Findings of the Association for Computational Linguistics: NAACL 2024}}.
\newblock


\bibitem[Ramos et~al\mbox{.}(2003)]%
        {ramos2003using}
\bibfield{author}{\bibinfo{person}{Juan Ramos} {et~al\mbox{.}}} \bibinfo{year}{2003}\natexlab{}.
\newblock \showarticletitle{Using tf-idf to determine word relevance in document queries}. In \bibinfo{booktitle}{\emph{Proceedings of the first instructional conference on machine learning}}, Vol.~\bibinfo{volume}{242}. Citeseer, \bibinfo{pages}{29--48}.
\newblock


\bibitem[Sarmah et~al\mbox{.}(2024)]%
        {sarmah2024hybridrag}
\bibfield{author}{\bibinfo{person}{Bhaskarjit Sarmah}, \bibinfo{person}{Benika Hall}, \bibinfo{person}{Rohan Rao}, \bibinfo{person}{Sunil Patel}, \bibinfo{person}{Stefano Pasquali}, {and} \bibinfo{person}{Dhagash Mehta}.} \bibinfo{year}{2024}\natexlab{}.
\newblock \showarticletitle{HybridRAG: Integrating Knowledge Graphs and Vector Retrieval Augmented Generation for Efficient Information Extraction}.
\newblock \bibinfo{journal}{\emph{arXiv preprint arXiv:2408.04948}} (\bibinfo{year}{2024}).
\newblock


\bibitem[Trivedi et~al\mbox{.}(2022)]%
        {trivedi2022musique}
\bibfield{author}{\bibinfo{person}{Harsh Trivedi}, \bibinfo{person}{Niranjan Balasubramanian}, \bibinfo{person}{Tushar Khot}, {and} \bibinfo{person}{Ashish Sabharwal}.} \bibinfo{year}{2022}\natexlab{}.
\newblock \showarticletitle{MuSiQue: Multihop Questions via Single-hop Question Composition}.
\newblock \bibinfo{journal}{\emph{Transactions of the Association for Computational Linguistics}}  \bibinfo{volume}{10} (\bibinfo{year}{2022}), \bibinfo{pages}{539--554}.
\newblock


\bibitem[Wang et~al\mbox{.}(2022)]%
        {wang2022rete}
\bibfield{author}{\bibinfo{person}{Ruijie Wang}, \bibinfo{person}{Zheng Li}, \bibinfo{person}{Danqing Zhang}, \bibinfo{person}{Qingyu Yin}, \bibinfo{person}{Tong Zhao}, \bibinfo{person}{Bing Yin}, {and} \bibinfo{person}{Tarek Abdelzaher}.} \bibinfo{year}{2022}\natexlab{}.
\newblock \showarticletitle{RETE: retrieval-enhanced temporal event forecasting on unified query product evolutionary graph}. In \bibinfo{booktitle}{\emph{Proceedings of the ACM Web Conference 2022}}. \bibinfo{pages}{462--472}.
\newblock


\bibitem[Wang et~al\mbox{.}(2024)]%
        {wang2024knowledge}
\bibfield{author}{\bibinfo{person}{Yu Wang}, \bibinfo{person}{Nedim Lipka}, \bibinfo{person}{Ryan~A Rossi}, \bibinfo{person}{Alexa Siu}, \bibinfo{person}{Ruiyi Zhang}, {and} \bibinfo{person}{Tyler Derr}.} \bibinfo{year}{2024}\natexlab{}.
\newblock \showarticletitle{Knowledge graph prompting for multi-document question answering}. In \bibinfo{booktitle}{\emph{Proceedings of the AAAI Conference on Artificial Intelligence}}, Vol.~\bibinfo{volume}{38}. \bibinfo{pages}{19206--19214}.
\newblock


\bibitem[Xu et~al\mbox{.}(2024)]%
        {xu2024retrieval}
\bibfield{author}{\bibinfo{person}{Zhentao Xu}, \bibinfo{person}{Mark~Jerome Cruz}, \bibinfo{person}{Matthew Guevara}, \bibinfo{person}{Tie Wang}, \bibinfo{person}{Manasi Deshpande}, \bibinfo{person}{Xiaofeng Wang}, {and} \bibinfo{person}{Zheng Li}.} \bibinfo{year}{2024}\natexlab{}.
\newblock \showarticletitle{Retrieval-augmented generation with knowledge graphs for customer service question answering}. In \bibinfo{booktitle}{\emph{Proceedings of the 47th International ACM SIGIR Conference on Research and Development in Information Retrieval}}. \bibinfo{pages}{2905--2909}.
\newblock


\bibitem[Yang et~al\mbox{.}(2018)]%
        {yang2018hotpotqa}
\bibfield{author}{\bibinfo{person}{Zhilin Yang}, \bibinfo{person}{Peng Qi}, \bibinfo{person}{Saizheng Zhang}, \bibinfo{person}{Yoshua Bengio}, \bibinfo{person}{William Cohen}, \bibinfo{person}{Ruslan Salakhutdinov}, {and} \bibinfo{person}{Christopher~D Manning}.} \bibinfo{year}{2018}\natexlab{}.
\newblock \showarticletitle{HotpotQA: A Dataset for Diverse, Explainable Multi-hop Question Answering}. In \bibinfo{booktitle}{\emph{Proceedings of the 2018 Conference on Empirical Methods in Natural Language Processing}}. \bibinfo{pages}{2369--2380}.
\newblock


\bibitem[Yu et~al\mbox{.}(2024)]%
        {yu2024evaluation}
\bibfield{author}{\bibinfo{person}{Hao Yu}, \bibinfo{person}{Aoran Gan}, \bibinfo{person}{Kai Zhang}, \bibinfo{person}{Shiwei Tong}, \bibinfo{person}{Qi Liu}, {and} \bibinfo{person}{Zhaofeng Liu}.} \bibinfo{year}{2024}\natexlab{}.
\newblock \showarticletitle{Evaluation of retrieval-augmented generation: A survey}. In \bibinfo{booktitle}{\emph{CCF Conference on Big Data}}. Springer, \bibinfo{pages}{102--120}.
\newblock


\bibitem[Zhou et~al\mbox{.}(2025)]%
        {zhou2025depth}
\bibfield{author}{\bibinfo{person}{Yingli Zhou}, \bibinfo{person}{Yaodong Su}, \bibinfo{person}{Youran Sun}, \bibinfo{person}{Shu Wang}, \bibinfo{person}{Taotao Wang}, \bibinfo{person}{Runyuan He}, \bibinfo{person}{Yongwei Zhang}, \bibinfo{person}{Sicong Liang}, \bibinfo{person}{Xilin Liu}, \bibinfo{person}{Yuchi Ma}, {et~al\mbox{.}}} \bibinfo{year}{2025}\natexlab{}.
\newblock \showarticletitle{In-depth Analysis of Graph-based RAG in a Unified Framework}.
\newblock \bibinfo{journal}{\emph{arXiv preprint arXiv:2503.04338}} (\bibinfo{year}{2025}).
\newblock


\end{thebibliography}
